 \documentclass[preprint,12pt]{elsarticle}%review,
\usepackage{amssymb,amsmath,latexsym}
\usepackage[dvips]{color}
\usepackage[headings]{fullpage}
\usepackage{epic,epsfig,graphicx}
\usepackage{enumerate}
\usepackage{multirow}
\usepackage{setspace}
\usepackage{keyval}
\usepackage{float}
\usepackage{amssymb}
 \newtheorem{thm}{Theorem}[section]

 \newtheorem{lemma}{Lemma}[section]

 \newtheorem{rem}{Remark}[section]
  \newtheorem{assumption}{Assumption}[section]

 \numberwithin{equation}{section}
\journal{***}

\begin{document}
\begin{frontmatter}
\title{A study of the persistence and extinction of malaria in a family of epidemic models}
%{A Scale-Structured Network Stochastic Epidemic dynamic model with varying Incubation Period}
\author{Divine Wanduku and Broderick Oluyede }
\address{Department of Mathematical Sciences,
Georgia Southern University, 65 Georgia Ave, Room 3042, Statesboro,
Georgia, 30460, U.S.A. E-mail:dwanduku@georgiasouthern.edu;wandukudivine@yahoo.com\footnote{Corresponding author. Tel: +14073009605.
} }%fax: +18139746720.
\begin{abstract}
 This paper investigates the deterministic extinction and permanence of a family of SEIRS malaria models with multiple random delays, and with a general nonlinear incidence rate. The conditions for the extinction and permanence of the disease are presented. For the persistence of disease, improved analytical techniques are employed to obtain eventual lower bounds for the states of the system, while extinction results are obtained via analyzing the Lyapunov exponent. The basic reproduction number $R^{*}_{0}$ is calculated, and for $R^{*}_{0}<1$, extinction of disease occurs, while for $R^{*}_{0}\geq 1$, the disease persists. But if $R^{*}_{0}\geq 1$, and the expected survival probability of the plasmodium   $E(e^{-(\mu_{v}T_{1}+\mu T_{2})})<\frac{1}{R^{*}_{0}}$, then extinction occurs. Results are interpreted, and numerical simulation examples are presented to (a) compare and evaluate the impacts of the intensity of the incidence rate, and the intensity of response to malaria control, on the persistence of malaria in the human population, and (2) to justify the asymptotic stability of the disease-free equilibrium.
 \end{abstract}

\end{frontmatter}
\section{Introduction\label{ch1.sec0}}
Continuing earlier discussions about malaria in Wanduku\cite{wanduku-biomath}, malaria has exhibited an increasing alarming high mortality rate between 2015 and 2016. In fact, the latest WHO-\textit{World Malaria Report 2017} \cite{WHO-new} estimates a total of 216 million cases of malaria from 91 countries in 2016, which constitutes a 5 million increase in the total malaria cases from the malaria statistics obtained previously in 2015. Moreover, the total death count was 445000, and sub-Saharan Africa accounts for 90\% of the total estimated malaria cases. This rising trend in the malaria data, signals a need for more learning about the disease, improvement of the existing control strategies and equipment, and also a need for more advanced  resources etc. to fight  and eradicate, or ameliorate the burdens of malaria.

 Malaria and other mosquito-borne diseases such as dengue fever, yellow fever, zika fever, lymphatic filariasis, and the different types of encephalitis etc. exhibit some  unique biological features. For instance, the incubation of the disease requires two hosts - the mosquito vector and human hosts,  which may be either  directly involved in a full life cycle of the infectious agent consisting of two separate and independent segments of sub-life cycles, which are completed separately inside the two hosts, or directly involved in two separate and independent half-life cycles of the infectious agent in the hosts. Therefore, there is a total latent time lapse of disease incubation which extends over the two segments of delay incubation times namely:- (1) the incubation period of the infectious agent ( or the half-life cycle) inside the vector, and (2) the incubation period of the infectious agent (or the other half-life cycle) inside the human being. See \cite{WHO,CDC}. In fact,  the malaria plasmodium undergoes the first developmental half-life cycle called the \textit{sporogonic cycle} inside the female \textit{Anopheles} mosquito lasting approximately $10-18$ days, following a successful blood meal obtained from an infectious human being through a mosquito bite. Moreover, the mosquito  becomes infectious.  The parasite  completes the second developmental half-life cycle called the \textit{exo-erythrocytic cycle} lasting about 7-30 days inside the exposed human being\cite{WHO,CDC}, whenever the parasite is transferred to  human being in the process of the infectious mosquito foraging for another blood meal.
%%%
 %Several vector-borne diseases  induce or confer natural immunity against the disease after infection and recovery. The effectiveness and duration of the natural protective immunity varies depending on the type of disease and also on other biological factors. For example, the exposure and successful recovery from one dengue fever viral strain confers lifelong immunity against the particular viral serotype\cite{WHO}.

The exposure and successful recovery from a malaria parasite, for example, \textit{falciparum vivae} induces natural immunity against the disease which can protect against subsequent severe outbreaks of the disease. Moreover, the effectiveness and duration of the naturally acquired immunity against malaria is determined by several factors such as the  species and the frequency of  exposure to the parasites. Furthermore, it has been determined that other biological factors such as the genetics of the human being, for instance, sickle-cell anaemia, duffy negative blood types have bearings on the naturally acquired immunity against different species of malaria\cite{CDC,lars,denise}.

 Compartmental mathematical epidemic dynamic models  have been used to investigate the dynamics of several different types of infectious diseases including malaria\cite{pang,gungala,hyun}. In general, these models are classified as SIS, SIR, SIRS, SEIRS,  and  SEIR etc.\cite{koro,mao-2,wanduku-fundamental,Wanduku-2017} epidemic dynamic models depending on the compartments of the disease classes directly involved in the general disease dynamics.
  Some of these studies devote interest to SEIRS and SEIR models\cite{sen,cooke-driessche,zhica}, which account for the compartment of individuals who are exposed to the disease, $E$, that is, infected but noninfectious individuals. This inclusion of the exposed class of individuals allows for more insights about the disease dynamics during the incubation stage of the disease.

 In addition, many of these epidemic dynamic models are improved in reality by including the time delays that occur in the disease dynamics. Generally, two distinct types of delays are studied namely:-disease latency and immunity delay. The disease latency represents the period of disease incubation, or period of infectiousness which nonetheless is studied as a delay in the disease dynamics. The immunity delay represents the period of effective naturally acquired immunity against the disease after successful recovery from infection. See \cite{Wanduku-2017,wanduku-delay,kyrychko,cooke,cooke-driessche}.

 Some important investigations in the study of population dynamic models expressed as  systems of differential equations are the persistence (or permanence) and extinction of disease in the population, and also disease eradication when the population is in a steady state over sufficiently long time. Several papers in the literature\cite{zhien,aadil,wanbiao,tend,mao-2,zhica}  have addressed these topics. The extinction of disease seeks to find conditions that are sufficient for the disease related classes in the population such as, the exposed and infectious classes, to become extinct over sufficiently long time. The persistence or permanence of disease also answers the question about whether a significant number of people in the disease related classes will remain over sufficiently long time.  Disease eradication in the steady state population seeks to find conditions sufficient for the disease-free equilibrium population to be stable asymptotically.

  The information about these three important properties of systems of differential equations is obtained via analyzing the behavior of the trajectories of the system over time in the neighborhood of the equilibria of the dynamic systems. Indeed, extinction of the disease conditions in the differential equation system  can be obtained by conducting a Lyapunov exponential analysis, and to determine convergence to the disease-free equilibrium. Persistence of disease conditions over time in the differential equation system can be obtained by analyzing the behavior of the solution of the system near an endemic steady state. And disease eradication conditions when the differential equation system is in steady state are obtained as sufficient conditions for every trajectory that starts in the neighborhood of the disease-free steady state to remain in the neighborhood of the steady state, and converge asymptotically to the steady state\cite{zhien,aadil,wanbiao,tend,mao-2,zhica}.

  The primary objectives of this paper include, to investigate (1) the extinction,  and (2) the persistence of malaria in a human population in the line of thinking of \cite{zhien}. In other words, we find conditions that are sufficient for the malaria parasite to become extinct from the population over time, and also conditions that would negatively prevail the disease in the population.
Recently, Wanduku\cite{wanduku-biomath} presented and studied the following novel family of epidemic dynamic models for malaria with three distributed delays:
%%%%%%%%%%%%%%%%
%

\begin{equation}\label{ch1.sec0.eq3}
\left\{
\begin{array}{lll}
dS(t)&=&\left[ B-\beta S(t)\int^{h_{1}}_{t_{0}}f_{T_{1}}(s) e^{-\mu s}G(I(t-s))ds - \mu S(t)+ \alpha \int_{t_{0}}^{\infty}f_{T_{3}}(r)I(t-r)e^{-\mu r}dr \right]dt,\\%\nonumber\\
%&&\label{ch1.sec0.eq3}\\
dE(t)&=& \left[ \beta S(t)\int^{h_{1}}_{t_{0}}f_{T_{1}}(s) e^{-\mu s}G(I(t-s))ds - \mu E(t)\right.\\%\nonumber\\
&&\left.-\beta \int_{t_{0}}^{h_{2}}f_{T_{2}}(u)S(t-u)\int^{h_{1}}_{t_{0}}f_{T_{1}}(s) e^{-\mu s-\mu u}G(I(t-s-u))dsdu \right]dt,\\%\label{ch1.sec0.eq4}\\
%&&\nonumber\\
dI(t)&=& \left[\beta \int_{t_{0}}^{h_{2}}f_{T_{2}}(u)S(t-u)\int^{h_{1}}_{t_{0}}f_{T_{1}}(s) e^{-\mu s-\mu u}G(I(t-s-u))dsdu- (\mu +d+ \alpha) I(t) \right]dt,\\%\nonumber\\
%&&\label{ch1.sec0.eq5}\\
dR(t)&=&\left[ \alpha I(t) - \mu R(t)- \alpha \int_{t_{0}}^{\infty}f_{T_{3}}(r)I(t-r)e^{-\mu s}dr \right]dt,%\label{ch1.sec0.eq6}
\end{array}
\right.
\end{equation}
where the initial conditions are given in the following: let $h= h_{1}+ h_{2}$ and define
\begin{eqnarray}
&&\left(S(t),E(t), I(t), R(t)\right)
=\left(\varphi_{1}(t),\varphi_{2}(t), \varphi_{3}(t),\varphi_{4}(t)\right), t\in (-\infty,t_{0}],\nonumber\\% t\in [t_{0}-h,t_{0}],\quad and\quad=
&&\varphi_{k}\in \mathcal{C}((-\infty,t_{0}],\mathbb{R}_{+}),\forall k=1,2,3,4,\quad \varphi_{k}(t_{0})>0,\forall k=1,2,3,4,\nonumber\\
 \label{ch1.sec0.eq06a}
\end{eqnarray}
where $\mathcal{C}((-\infty,t_{0}],\mathbb{R}_{+})$ is the space of continuous functions with  the supremum norm
\begin{equation}\label{ch1.sec0.eq06b}
||\varphi||_{\infty}=\sup_{ t\leq t_{0}}{|\varphi(t)|}.
\end{equation}

The disease spreads in the human population of total size $ N(t)=S(t)+ E(t)+ I(t)+ R(t)$, where $S(t)$, $E(t)$, $I(t)$ and $R(t)$ represent the susceptible, exposed, infectious and naturally acquired immunity classes at time $t$, respectively. The positive constants $B$, and $\mu$ represent the constant birth and natural death rates, respectively. Furthermore, the disease related deathrate is denoted $d$. For simplicity the vector and human natural death rates are the same, and $\beta$ is the average effective contact rate per infected mosquito per unit time. The recovery rate from malaria with acquired immunity is $\alpha$. Also, the incubation delays inside the mosquito and human hosts are denoted $T_{1}$ and $T_{2}$, respectively, and the period of effective naturally acquired immunity is denoted $T_{3}$. Moreover, the delays are random variables with arbitrary densities denoted $f_{T_{1}}$, $f_{T_{2}}$ and $f_{T_{3}}$, and their supports given as $T_{1}\in[t_{0},h_{1}]$, $T_{2}\in[t_{0},h_{1}]$ and $T_{3}\in[t_{0}, +\infty)$.
%%%%
The nonlinear incidence function $G$ which signifies the response to disease transmission by the susceptible class as malaria increases in the population, satisfies the following assumptions  \begin{assumption}\label{ch1.sec0.assum1}
\begin{enumerate}
  \item [$A1$]$G(0)=0$; $A2$: $G(I)$ is strictly monotonic on $[0,\infty)$; $A3$: $G''(I)<0$;% $\Leftrightarrow$ $G(I)$ is differentiable concave on $[0,\infty)$.
   $A4$. $\lim_{I\rightarrow \infty}G(I)=C, 0\leq C<\infty$; % $\Leftrightarrow$  $G(I)$ has a horizontal asymptote $0\leq C<\infty$.
  and  $A5$: $G(I)\leq I, \forall I>0$.% $\Leftrightarrow$ $G(I)$ is at most as large as the identity function $f:I\mapsto I$ over the positive all $I\in (0,\infty)$.
\end{enumerate}
\end{assumption}
 %%%%%%%%
%%%%%%%% that characterize the nonlinear behavior of the  incidence function $G(I)$.
More details about the derivation of the model in (\ref{ch1.sec0.eq3}) is given in Wanduku\cite{wanduku-biomath}. Whilst permanence of diseases in some delay type systems are known, such as for single finite, or single distributed, and also for some double finite delay systems ( cf.\cite{zhien, wanbiao,tend, zhica}), the permanence of  disease in systems with multiple random delays that occur in series\footnote{Delays that occur in series in this write-up have a superimposed effect.} is not properly understood in the literature. It appears that only these papers \cite{wanduku-biomath,shuj} have addressed some properties of differential equation systems with multiple delays in series.
Furthermore,  as far as we know no other paper has addressed extinction and persistence of malaria in a human population by conducting a Lyapunov exponent analysis, or analyzing the behavior of the trajectories of the differential equation system in the mode of thinking and techniques in \cite{zhien}.

This paper also presents an inherent algorithmic technique to analyze the permanence of diseases in  complex multiple distributed delay systems in the line of thinking of \cite{wanbiao}. In this regard, the primary goal is to add to the literature an analytic algorithmic method with numerical justification, to investigate deterministic permanence of complex multiple random delay systems.
%%%%%%%%(\ref{ch1.sec0.eq9}) and (\ref{ch1.sec0.eq11}), and the corresponding equations

The rest of this paper is presented as follows:- in Section~\ref{ch1.sec1}, some preliminary results for (\ref{ch1.sec0.eq3}) are presented. In Section~\ref{ch1.sec3a}, the results for the permanence of the disease are presented. Moreover, simulation results for the permanence of the disease in the population are presented in Section~\ref{ch1.sec4}. In Section~\ref{ch1.sec2b}, the results for the extinction of the disease are presented. Moreover, the numerical simulation results for the extinction of disease are presented in Section~\ref{ch1.sec4}.

Observe from (\ref{ch1.sec0.eq3}) that the equations for $E$ and $R$ decouple from the other two equations in the  system. Therefore, the results in this paper will be shown for the decoupled system containing equations for $S$ and $I$.
Nevertheless, the following notations are utilized:
\begin{equation}\label{ch1.sec0.eq13b}
%\left\{
%  \begin{array}{lll}
    Y(t)=(S(t), E(t), I(t), R(t))^{T},X(t)=(S(t),E(t),I(t))^{T},\quad\textrm{and}\quad N(t)=S(t)+ E(t)+ I(t)+ R(t).
  %\end{array}
%  \right.
\end{equation}
\section{Model Validation and Preliminary Results\label{ch1.sec1}}
The results in [Theorem~3.1, Wanduku\cite{wanduku-biomath}] show that the system (\ref{ch1.sec0.eq3}) has a unique positive solution $Y(t)\in \mathbb{R}_{+}^{4}$. Moreover,
\begin{equation}\label{ch1.sec1.thm1a.eq0}
\limsup_{t\rightarrow \infty} N(t)\leq S^{*}_{0}=\frac{B}{\mu}.
   \end{equation}
  Furthermore, there is a positive self invariant space for the system denoted $D(\infty)=\bar{B}^{(-\infty, \infty)}_{\mathbb{R}^{4}_{+},}\left(0,\frac{B}{\mu}\right) $, where $D(\infty)$ is the closed ball in $\mathbb{R}^{4}_{+}$ centered at the origin with radius $\frac{B}{\mu}$ containing all positive solutions defined over $(-\infty,\infty )$.
  % as
  %%%%%%%%%%%%%%%

In the analysis of the deterministic malaria model (\ref{ch1.sec0.eq3}) with initial conditions in (\ref{ch1.sec0.eq06a})-(\ref{ch1.sec0.eq06b})  in Wanduku\cite{wanduku-biomath}, the threshold values for disease eradication such as the basic reproduction number for the disease when the system is in steady state are obtained in both cases where the delays in the system $T_{1}, T_{2}$ and $T_{3}$ are constant, and  also arbitrarily distributed.

For $S^{*}_{0}=\frac{B}{\mu}$, when the delays in the system are all constant, the basic reproduction number of the disease is given by
\begin{equation} \label{ch1.sec2.lemma2a.corrolary1.eq4}
\hat{R}^{*}_{0}=\frac{\beta S^{*}_{0} }{(\mu+d+\alpha)}.
\end{equation}
Furthermore, the threshold condition $\hat{R}^{*}_{0}<1$ is required for the disease-free equilibrium $E_{0}=X^{*}_{0}=(S^{*}_{0},0,0)$ to be asymptotically stable, and hence, for the disease to be eradicated from the steady state human population.

On the other hand, when the delays in the system  $T_{i}, i=1,2,3$  are random, and arbitrarily distributed, the basic reproduction number is given by
\begin{equation}\label{ch1.sec2.theorem1.corollary1.eq3}
R_{0}=\frac{\beta S^{*}_{0} \hat{K}_{0}}{(\mu+d+\alpha)}+\frac{\alpha}{(\mu+d+\alpha)},
\end{equation}
where, $\hat{K}_{0}>0$ is a constant that depends only on $S^{*}_{0}$  (in fact, $\hat{K}_{0}=4+ S^{*}_{0} $). In addition, malaria is eradicated from the system in the steady state, whenever $R_{0}\leq 1$,
%%%%%%%%%%%%%%%%%%
%%%%%%%%%%%%%%%%%%%%%%%

 The results in [Theorem~5.1, Wanduku\cite{wanduku-biomath}] also show that when $R_{0}>1$, and the delays in the system $T_{i}, i=1, 2, 3$ are random, and arbitrarily distributed, the deterministic system (\ref{ch1.sec0.eq3}) has a unique positive equilibrium state denoted by $E_{1}=(S^{*}_{1}, E^{*}_{1}, I^{*}_{1})$.
%%%%
%  %%%%%%%%%%%%%%
%  %%%%%%%%%%%%%
 %%%%%%%%%%%%%%%%%%%%%%%%%%%%
 \section{Permanence of the disease in the deterministic system\label{ch1.sec3a}}
 %%%%%%%%%%%%
 %%%%%%%%%%%%%%%%%%%%%%%%%%%%
 The following lemma will be used to establish the results about the permanence of the disease in the population.% described by the system of differential equations
 \begin{lemma}\label{ch1.sec4.lemma1}
   Suppose the conditions of [Theorem~5.1, Wanduku\cite{wanduku-biomath}]%Theorem~\ref{ch1.sec3.thm1}
    are satisfied, and let the nonlinear incidence function $G$ characterized by the assumptions in Assumption~\ref{ch1.sec0.assum1}  satisfy the additional condition
   \begin{equation}\label{ch1.sec4.lemma1.eq1a}
     \left(\frac{G(x)}{x}-\frac{G(y)}{y}\right)\left(G(x)-G(y)\right)\leq 0, \forall x, y\geq0.
   \end{equation}
   Then every positive solution $(S(t), I(t))$ of  the  decoupled deterministic system (\ref{ch1.sec0.eq3}) with initial conditions (\ref{ch1.sec0.eq06a}) and (\ref{ch1.sec0.eq06b}) satisfies the following conditions:
    \begin{equation}\label{ch1.sec4.lemma1.eq1}
    \liminf_{t\rightarrow \infty}{S(t)}\geq v_{1}\equiv \frac{B}{\mu+\beta G(S_{0})}\quad and\quad  \liminf_{t\rightarrow \infty}{I(t)}\geq v_{2}\equiv qI^{*}_{1}e^{-(\mu+d+\alpha)(\rho+1)h},
  \end{equation}
  where $h=h_{1}+h_{2}$,  and  $\rho>0$ is a suitable positive constant,  $S^{*}_{1}<\min\{S_{0}, S^{\vartriangle}\}$ and $0<q<\bar{q}<1$, given that,
  \begin{equation}\label{ch1.sec4.lemma1.eq2}
  \bar{q}=\frac{B\beta E(e^{-\mu T_{1}})G(I^{*}_{1})-\mu \alpha E(e^{-\mu T_{3}})I^{*}_{1}}{\left(B+\alpha E(e^{-\mu T_{3}})I^{*}_{1}\right)\beta I^{*}_{1}},\quad S^{\vartriangle}=\frac{B}{k}\left(1-e^{-k\rho h}\right), k=\mu +\beta G(q I^{*}_{1}).
  \end{equation}
 \end{lemma}
 %%%%%%%%%%%%%%%%%%%%%%%%%%
 %%%%%%%%%%%%%%%%%%%%[Theorem~3.1, Wanduku\cite{wanduku-biomath}] show that the system (\ref{ch1.sec0.eq3}) has a positive solution $Y(t)\in \mathbb{R}_{+}^{4}$. Moreover,
%\begin{equation}\label{ch1.sec1.thm1a.eq0}Theorem~\ref{ch1.sec1.thm1a}
 Proof:\\
 Recall, [Theorem~3.1, Wanduku\cite{wanduku-biomath}] and (\ref{ch1.sec1.thm1a.eq0}) assert that for $N(t)=S(t)+E(t)+ I(t)+R(t)$, $\limsup_{t\rightarrow \infty} {N(t)}\leq S^{*}_{0}=\frac{B}{\mu}$. This implies that  $\limsup_{t\rightarrow \infty} {S(t)}\leq S^{*}_{0}$. This further implies that for any arbitrarily small $\epsilon>0$, there exists a sufficiently large $\Lambda>0$, such that
 \begin{equation}\label{ch1.sec4.lemma1.proof.eq1}
 I(t)\leq S^{*}_{0} +\varepsilon,\quad whenever, \quad t\geq \Lambda.
 \end{equation}
 Without loss of generality, let $\Lambda_{1}>0$ be sufficiently large such that
 \[t\geq \Lambda\geq \max_{(s, r)\in [t_{0}, h_{1}]\times [t_{0}, \infty)}{(\Lambda_{1}+s, \Lambda_{1}+r)}.\]
   It follows from Assumption~\ref{ch1.sec0.assum1}  and (\ref{ch1.sec0.eq3}) that
 \begin{eqnarray}
 \frac{dS(t)}{dt}&\geq& B-\beta S(t)\int_{t_{0}}^{h_{1}}f_{T_{1}}(s)e^{\mu s}G(S^{*}_{0}+\epsilon)ds-\mu S(t),\nonumber\\
 &\geq& B-\left[\mu+ \beta G(S^{*}_{0}+\epsilon)\right]S(t).\label{ch1.sec4.lemma1.proof.eq2}
 \end{eqnarray}
 From (\ref{ch1.sec4.lemma1.proof.eq2}) it follows that
 \begin{equation}\label{ch1.sec4.lemma1.proof.eq3}
 S(t)\geq  \frac{B}{k_{1}} -\frac{B}{k_{1}}e^{-k_{1}(t-t_{0})}+S(t_{0}) e^{-k_{1}(t-t_{0})},
 \end{equation}
 where $k_{1}=\mu+ \beta G(S^{*}_{0}+\epsilon)$.

 It is easy to see from (\ref{ch1.sec4.lemma1.proof.eq3})
 \begin{equation}\label{ch1.sec4.lemma1.proof.eq4}
   \liminf_{t\rightarrow \infty}{S(t)}\geq  \frac{B}{\mu+\beta G(S_{0}+\epsilon)}.
 \end{equation}
 Since $\epsilon>0$ is arbitrarily small, then the first part of (\ref{ch1.sec4.lemma1.eq1}) follows immediately.

 In the following it is shown that $ \liminf_{t\rightarrow \infty}{I(t)}\geq v_{2}$. In order to establish this result, it is first proved that it is impossible that $I(t)\leq q I^{*}_{1}$ for sufficiently large  $t\geq t_{0}$, where $q\in(0, 1)$ is defined in the hypothesis. Suppose on the contrary there exists some sufficiently large $\Lambda_{0}> t_{0}>0$, such that $I(t)\leq q I^{*}_{1}, \forall t\geq \Lambda_{0}$. It follows from (\ref{ch1.sec0.eq3}) that
 \begin{eqnarray}
   S^{*}_{1} &=& \frac{B+\alpha E(e^{-\mu T_{3}})I^{*}_{1}}{\mu+ \beta E(e^{-\mu T_{1}})G(I^{*}_{1})} \nonumber\\
    &=& \frac{B}{\mu+\frac{B\beta E(e^{-\mu T_{1}})G(I^{*}_{1})-\mu \alpha E(e^{-\mu T_{3}})I^{*}_{1} }{B+\alpha E(e^{-\mu T_{3}})I^{*}_{1}}}.\label{ch1.sec4.lemma1.proof.eq5}
 \end{eqnarray}
 But, it can be easily seen from (\ref{ch1.sec0.eq3}) that
 \begin{eqnarray}
    B\beta E(e^{-\mu T_{1}})G(I^{*}_{1})-\mu \alpha E(e^{-\mu T_{3}})I^{*}_{1}&=&\frac{\mu(\mu+d+\alpha)\left[S^{*}_{0}-\frac{\alpha E(e^{-\mu T_{3}})E(e^{-\mu T_{2}})}{(\mu +d+\alpha)}S^{*}_{1}\right]}{E(e^{-\mu T_{2}})S^{*}_{1}}I^{*}_{1}\nonumber \\
   &\geq& \frac{\mu(\mu+d+\alpha)(S^{*}_{0}-S^{*}_{1})}{E(e^{-\mu T_{2}})S^{*}_{1}}\nonumber\\
   &>&0,\quad since\quad S^{*}_{0}=\frac{B}{\mu}\geq S^{*}_{1}.\label{ch1.sec4.lemma1.proof.eq6}
    \end{eqnarray}
   Therefore, from (\ref{ch1.sec4.lemma1.proof.eq5}), it follows that
   \begin{equation}\label{ch1.sec4.lemma1.proof.eq6}
     S^{*}_{1}<\frac{B}{\mu +\beta I^{*}_{1}q}\leq \frac{B}{\mu +\beta G(qI^{*}_{1})},
   \end{equation}
   where $0<q<\bar{q}$, and $\bar{q}$ is defined in (\ref{ch1.sec4.lemma1.eq2}).

   For all  vector values $ (s, r)\in [t_{0}, h_{1}]\times [t_{0}, \infty) $ define
   \begin{equation}
   \Lambda_{0,max}=\max_{(s, r)\in [t_{0}, h_{1}]\times [t_{0}, \infty)}{(\Lambda_{0}+s, \Lambda_{0}+r)},
   \end{equation}
It follows from Assumption~\ref{ch1.sec0.assum1}  and (\ref{ch1.sec0.eq3}) that for all $t\geq \Lambda_{0,max}$,
\begin{equation}\label{ch1.sec4.lemma1.proof.eq7}
   S(t)\geq  \frac{B}{k} -\frac{B}{k}e^{-k(t-\Lambda_{0,max})}+S(\Lambda_{0,max}) e^{-k(t-\Lambda_{0,max})},
\end{equation}
where $k$ is defined in (\ref{ch1.sec4.lemma1.eq2}).
 For $t\geq \Lambda_{0,max}+ \rho h$,  where $ h=h_{1}+h_{2}$, and $\rho>0$ is sufficiently large,  it follows from (\ref{ch1.sec4.lemma1.proof.eq7}) that
 \begin{equation}\label{ch1.sec4.lemma1.proof.eq8}
   S(t)\geq \frac{B}{k}\left[1-e^{-k(t-\Lambda_{0,max})}\right]\geq \frac{B}{k}\left[1-e^{-k\rho h}\right]=S^{\vartriangle}.
 \end{equation}
 Hence, from (\ref{ch1.sec4.lemma1.proof.eq6}) and (\ref{ch1.sec4.lemma1.proof.eq8}), it follows that for some suitable choice of $\rho>0$ sufficiently large, then
 \begin{equation}\label{ch1.sec4.lemma1.proof.eq9}
   S^{\vartriangle}>S^{*}_{1}, \forall t\geq \Lambda_{0,max}+ \rho h.
 \end{equation}

 For $t\geq \Lambda_{0,max}+ \rho h$, define
 \begin{eqnarray}
   V(t) &=& I(t) + \beta S^{*}_{1}\int_{t_{0}}^{h_{2}}\int_{t_{0}}^{h_{1}}f_{T_{2}}(u)f_{T_{1}}(s)e^{-\mu(s+u)}\int_{t-s}^{t}G(I(v-u))dvdsdu \nonumber\\
    &&+ \beta S^{*}_{1}\int_{t_{0}}^{h_{2}}\int_{t_{0}}^{h_{1}}f_{T_{2}}(u)f_{T_{1}}(s)e^{-\mu(s+u)}\int_{t-u}^{t}G(I(v))dvdsdu. \label{ch1.sec4.lemma1.proof.eq10}
 \end{eqnarray}
 It is easy to see from (\ref{ch1.sec0.eq3}) and (\ref{ch1.sec4.lemma1.proof.eq10}) that differentiating $V(t)$ with respect to the system (\ref{ch1.sec0.eq3}), leads to the following
 \begin{eqnarray}
   \dot{V}(t) &=&\beta \int_{t_{0}}^{h_{2}}\int_{t_{0}}^{h_{1}}f_{T_{2}}(u)f_{T_{1}}(s)e^{-\mu(s+u)}G(I(t-s-u))[S(t-u)-S^{*}_{1}]dsdu  \nonumber\\
   &&+\left[\beta S^{*}_{1}E(e^{-\mu(T_{1}+T_{2})})\frac{G(I(t))}{I(t)}-(\mu + d+\alpha)\right]I(t).\label{ch1.sec4.lemma1.proof.eq11}
 \end{eqnarray}
 For all $t\geq \Lambda_{0,max}+ \rho h +h>\Lambda_{0,max}+ \rho h +h_{2}$, it follows from (\ref{ch1.sec4.lemma1.eq1a}), (\ref{ch1.sec4.lemma1.proof.eq9}) and (\ref{ch1.sec0.eq3}) that
 \begin{eqnarray}
   \dot{V}(t) &\geq&\beta \int_{t_{0}}^{h_{2}}\int_{t_{0}}^{h_{1}}f_{T_{2}}(u)f_{T_{1}}(s)e^{-\mu(s+u)}G(I(t-s-u))[S^{\vartriangle}-S^{*}_{1}]dsdu \nonumber \\
   &&+\left[\beta S^{*}_{1}E(e^{-\mu(T_{1}+T_{2})})\frac{G(I^{*}_{1})}{I^{*}_{1}}-(\mu + d+\alpha)\right]I(t)\nonumber \\
   &=&\beta \int_{t_{0}}^{h_{2}}\int_{t_{0}}^{h_{1}}f_{T_{2}}(u)f_{T_{1}}(s)e^{-\mu(s+u)}G(I(t-s-u))[S^{\vartriangle}-S^{*}_{1}]dsdu.\label{ch1.sec4.lemma1.proof.eq11b}
 \end{eqnarray}
 Observe that the union of the subintervals  $\bigcup _{(s, u)\in [t_{0}, h_{1}]\times[t_{0}, h_{2}]}{[t_{0}-(s+u), t_{0}]}=[t_{0}-h, t_{0}]$, where $ h=h_{1}+h_{2}$. Denote the following
 \begin{equation}\label{ch1.sec4.lemma1.proof.eq11c}
   i_{min}=\min_{\theta\in [t_{0}-h, t_{0}], (s, u)\in [t_{0}, h_{1}]\times[t_{0}, h_{2}]} {I(\Lambda_{0,max}+ \rho h+ h+s+u+\theta)}.
 \end{equation}
 Note that (\ref{ch1.sec4.lemma1.proof.eq11c}) is equivalent to
 \begin{equation}\label{ch1.sec4.lemma1.proof.eq11a}
   i_{min}=\min_{\theta\in [t_{0}-h, t_{0}]} {I(\Lambda_{0,max}+ \rho h+ h+h+\theta)}.
 \end{equation}
 It is shown in the following that $I(t)\geq i_{min}, \forall t\geq \Lambda_{0,max}+ \rho h+h\geq \Lambda_{0,max}+ \rho h+u$, $\forall u\in [t_{0}, h_{2}]$.

 Suppose on the contrary there exists $\tau_{1}\geq 0$ such that  $I(t)\geq i_{min}$ for all
  $t\in [\Lambda_{0,max}+ \rho h +h, \Lambda_{0,max}+ \rho h +h+h+\tau_{1}]\supset [\Lambda_{0,max}+ \rho h +h, \Lambda_{0,max}+ \rho h +h+ s+u+\tau_{1}], \forall (s, u)\in [t_{0}, h_{1}]\times[t_{0}, h_{2}]$
  %%%%%%%%%%
  %$\Lambda_{0,max}+ \rho h +h\leq t\leq \Lambda_{0,max}+ \rho h +h+ s+u+\tau_{1}\leq \Lambda_{0,max}+ \rho h +h_{2}+h+\tau_{1}, \forall (s, u)\in [t_{0}, h_{1}]\times[t_{0}, h_{2}]$,
  %and the two variable functions
 %%\leq \Lambda_{0,max}+ \rho h +2u+ s+\tau_{1}
 \begin{equation}\label{ch1.sec4.lemma1.proof.eq12}
   I(\Lambda_{0,max}+ \rho h +h+h+\tau_{1})=i_{min},\quad and\quad  \dot{I}(\Lambda_{0,max}+ \rho h +h+h+\tau_{1})\leq 0.%\quad \forall  (s, u)\in [t_{0}, h_{1}]\times[t_{0}, h_{2}].
 \end{equation}
 For the value of $t=\Lambda_{0,max}+ \rho h +h+h+\tau_{1}$, it follows that $ S(t-u)>S^{\vartriangle}>S^{*}_{1}$, and $t-s-u\in [\Lambda_{0,max}+ \rho h +h, \Lambda_{0,max}+ \rho h +h+h+\tau_{1}]$, $ \forall  (s, u)\in [t_{0}, h_{1}]\times[t_{0}, h_{2}]$, and  it can be further seen from (\ref{ch1.sec0.eq3}), (\ref{ch1.sec4.lemma1.proof.eq9}) and (\ref{ch1.sec4.lemma1.eq1a}) that
 \begin{eqnarray}
   \dot{I}(t)&\geq& \beta E(e^{-\mu(T_{1}+T_{2})})G(i_{min})S^{\vartriangle}-(\mu+d+\alpha)i_{min},\nonumber \\
    &=&\left[\beta E(e^{-\mu(T_{1}+T_{2})})\frac{G(i_{min})}{i_{min}}S^{\vartriangle}-(\mu+d+\alpha)\right]i_{min}, \nonumber \\
    &>&\left[\beta E(e^{-\mu(T_{1}+T_{2})})\frac{G(I^{*}_{1})}{I^{*}_{1}}S^{*}_{1}-(\mu+d+\alpha)\right]i_{min}, \nonumber \\
    &=&0.\label{ch1.sec4.lemma1.proof.eq13}
 \end{eqnarray}
But (\ref{ch1.sec4.lemma1.proof.eq13}) contradicts (\ref{ch1.sec4.lemma1.proof.eq12}). Therefore, $I(t)\geq i_{min}, \forall t\geq \Lambda_{0,max}+ \rho h+h\geq \Lambda_{0,max}+ \rho h+u+s$, $\forall  (s, u)\in [t_{0}, h_{1}]\times[t_{0}, h_{2}]$.

 It follows further  from (\ref{ch1.sec4.lemma1.proof.eq11})-(\ref{ch1.sec4.lemma1.proof.eq11c}), and the Assumption~\ref{ch1.sec0.assum1} that for $\forall t\geq \Lambda_{0,max}+ \rho h+h+h\geq \Lambda_{0,max}+ \rho h+h+s+u$, $\forall (s,u)\in [t_{0}, h_{1}]\times[t_{0}, h_{2}]$.
 \begin{eqnarray}
   \dot{V}(t) &\geq&\beta \int_{t_{0}}^{h_{2}}\int_{t_{0}}^{h_{1}}f_{T_{2}}(u)f_{T_{1}}(s)e^{-\mu(s+u)}G(I(t-s-u))[S^{\vartriangle}-S^{*}_{1}]dsdu\nonumber\\
   &>& \beta E(e^{-\mu(T_{1}+T_{2})})G(i_{min})(S^{\vartriangle}-S^{*}_{1})>0.\label{ch1.sec4.lemma1.proof.eq14}
 \end{eqnarray}
 From (\ref{ch1.sec4.lemma1.proof.eq14}), it implies that $\limsup_{t\rightarrow\infty}{V(t)}=+\infty$.

  On the contrary, it can be seen from [Theorem~3.1, Wanduku\cite{wanduku-biomath}] and (\ref{ch1.sec1.thm1a.eq0}) that  $\limsup_{t\rightarrow \infty} N(t)\leq S^{*}_{0}=\frac{B}{\mu}$, which implies that $\limsup_{t\rightarrow \infty} I(t)\leq S^{*}_{0}=\frac{B}{\mu}$. This further implies that for every $\epsilon>0$ infinitesimally small, there exists $\tau_{2}>0$ sufficiently large such that $I(t)\leq S^{*}_{0}+\varepsilon, \forall t\geq \tau_{2}$.  It follows that from Assumption~\ref{ch1.sec0.assum1} that
  \begin{equation}\label{ch1.sec4.lemma1.proof.eq15}
    G(I(t-s-u))\leq G(I(v-u))\leq G(I(t-u))\leq G(I(t))\leq G(S^{*}_{0}+\epsilon), \forall v\in [t-s,t], (s,u)\in [t_{0}, h_{1}]\times[t_{0}, h_{2}].
  \end{equation}
  From (\ref{ch1.sec4.lemma1.proof.eq15}), it follows that
  \begin{equation}\label{ch1.sec4.lemma1.proof.eq16}
  \limsup_{t\rightarrow\infty}{G(I(t-s-u))}\leq \limsup_{t\rightarrow\infty}{G(I(t))}\leq G(S^{*}_{0}).
  \end{equation}
  It is easy to see from (\ref{ch1.sec4.lemma1.proof.eq10}) and (\ref{ch1.sec4.lemma1.proof.eq16}) that
  \begin{equation}\label{ch1.sec4.lemma1.proof.eq17}
    \limsup_{t\rightarrow\infty}{V(t)}\leq S^{*}_{0}+\beta S^{*}_{1}G(S^{*}_{0})E\left((T_{1}+T_{2})e^{-\mu(T_{1}+T_{2})}\right)<\infty.
  \end{equation}
  Therefore, it is impossible that $I(t)\leq q I^{*}_{1}$ for sufficiently large  $t\geq t_{0}$, where $q\in(0, 1)$.
  %%%%%%%%%%%%%%%%################

  Hence, the following are possible, $(Case(i.))$ $I(t)\geq qI^{*}_{1}$ for all $t$ sufficiently large, and $(Case(ii.))$ $I(t)$ oscillates about $qI^{*}_{1}$  for sufficiently large $t$. Obviously, we need show only $Case(ii.)$. Suppose $t_{1}$ and $t_{2}$ are are sufficiently large values such that
  \begin{equation}\label{ch1.sec4.lemma1.proof.eq18}
    I(t_{1})=I(t_{2})= qI^{*}_{1},\quad and\quad I(t)<qI^{*}_{1}, \forall (t_{1}, t_{2}).
  \end{equation}
  If for all $(s,u)\in [t_{0}, h_{1}]\times[t_{0}, h_{2}]$,  $t_{2}-t_{1}\leq \rho h+h$, where $ h=h_{1}+h_{2}$, observe that $[t_{1}, t_{1}+\rho h+s+u]\subseteq [t_{1}, t_{1}+\rho h+h]$, and  it is easy to see from (\ref{ch1.sec0.eq3}) by integration that
  \begin{equation}\label{ch1.sec4.lemma1.proof.eq19}
    I(t)\geq I(t_{1})e^{-(\mu+d+\alpha)(t-t_{1})}\geq qI^{*}_{1} e^{-(\mu+d+\alpha)(\rho+1)h}\equiv v_{2}.
  \end{equation}
 If for all $(s,u)\in [t_{0}, h_{1}]\times[t_{0}, h_{2}]$,  $t_{2}-t_{1}>\rho h+h\geq \rho h+ s+u$, then it can be seen easily that $I(t)\geq v_{2}$, for all $t\in [t_{1}, t_{1}+\rho h+s+u]\subseteq  [t_{1}, t_{1}+\rho h+h]$.

 Now, for each $t\in (\rho h+h, t_{2})\supseteq (\rho h+s+u, t_{2})$, $\forall (s,u)\in [t_{0}, h_{1}]\times[t_{0}, h_{2}]$,
  %$t_{1}+s+u\leq t_{1}+h\leq t< t_{2}$
  one can also claim that $I(t)\geq v_{2}$.  Indeed, as similarly shown above, suppose  on the  contrary for all $(s,u)\in [t_{0}, h_{1}]\times[t_{0}, h_{2}]$, $\exists T^{*}>0$ such that $I(t)\geq v_{2}$, $\forall t\in [t_{1}, t_{1}+\rho h+h+T^{*}]\supseteq [t_{1}, t_{1}+\rho h+s+u+T^{*}]$
   %for $t_{1}\leq t\leq t_{1}+s+u+T^{*}\leq t_{1}+h+T^{*}$
 \begin{equation}\label{ch1.sec4.lemma1.proof.eq20}
   I(t_{1}+\rho h+h+T^{*})=v_{2},\quad but \quad \dot{I}(t_{1}+\rho h+h+T^{*})\leq 0.
 \end{equation}
 It follows from (\ref{ch1.sec0.eq3}) and (\ref{ch1.sec4.lemma1.eq1a}) that for the value of $t=t_{1}+\rho h+h+T^{*}$, %$\forall (s,u)\in [t_{0}, h_{1}]\times[t_{0}, h_{2}]$,
 \begin{eqnarray}
   I(t) &\geq& \beta E(e^{-\mu(T_{1}+T_{2})})G(v_{2})S^{\vartriangle}-(\mu+d+\alpha)v_{2}\nonumber \\
    &>&\left[\beta E(e^{-\mu(T_{1}+T_{2})})\frac{G(v_{2})}{v_{2}}S^{*}_{1}-(\mu+d+\alpha)\right]v_{2}, \nonumber \\
    &\geq&\left[\beta E(e^{-\mu(T_{1}+T_{2})})\frac{G(I^{*}_{1})}{I^{*}_{1}}S^{*}_{1}-(\mu+d+\alpha)\right]v_{2}, \nonumber \\
    &=&0.\label{ch1.sec4.lemma1.proof.eq21}
 \end{eqnarray}
  Observe that (\ref{ch1.sec4.lemma1.proof.eq21}) contradicts (\ref{ch1.sec4.lemma1.proof.eq20}). Therefore, $I(t)\geq v_{2}$,  for $t\in [t_{1}, t_{2}]$.
  And since  $[t_{1}, t_{2}]$ is arbitrary, it implies that $I(t)\geq v_{2}$ for all sufficiently large $t$.  Therefore (\ref{ch1.sec4.lemma1.eq1}) is satisfied.
%
% From Lemma~\ref{ch1.sec4.lemma1}, the following theorem about the permanent of the disease in the population represented by the epidemic dynamic model (\ref{ch1.sec0.eq3})-(\ref{ch1.sec0.eq6}) is stated.
 \begin{thm}\label{ch1.sec4.thm1}
 If the conditions of Lemma~ \ref{ch1.sec4.lemma1} are satisfied, then the system (\ref{ch1.sec0.eq3}) is permanent for any total delay time $h=h_{1}+h_{2}$ (see \cite{zhien} for the definition of persistence of a state).
 \end{thm}
  \begin{rem}\label{ch1.sec4.rem1}
  \item[1.] It can be seen from Lemma~ \ref{ch1.sec4.lemma1} (\ref{ch1.sec4.lemma1.eq1}) that when $\beta=0$, then $v_{1}=\frac{B}{\mu}$. That is, when disease transmission stops, then asymptotically, the smallest total susceptible that remains will be all new births over the average lifespan of a human being in the population, which is also the disease free state $S^{*}_{0}=\frac{B}{\mu}$ (see \cite{wanduku-biomath}). Also, as the disease transmission rate rises given by $\beta\rightarrow\infty$, then the total susceptible that remains $v_{1}\rightarrow 0^{+}$. That is, as disease transmission rises, even the new births are either infected, or die from natural or disease related causes over sufficiently long time leaving only infinitesimally small number of susceptible people. These facts are exhibited in the example presented in the next section.
    \item[2.] From (\ref{ch1.sec4.lemma1.eq1}), observe that $e^{-(\mu+d+\alpha)(\rho+1)h}$ is the survival probability  from natural death ($\mu$), disease mortality ($d$), and from infectiousness ($\alpha$), over the total duration of the life cycle of the parasite $h$. Thus, the smallest total infectious class  that remains asymptotically $v_{2}\equiv qI^{*}_{1}e^{-(\mu+d+\alpha)(\rho+1)h}$ is a fraction $q\in(0,1)$ of the endemic equilibrium population $I^{*}_{1}$ that survives from all sources of death and disease over the parasite life cycle.

        Since $\frac{1}{(\mu+d+\alpha)}$ is the effective average lifespan of an individual who gets infected by malaria and survives the disease until recovery at rate $\alpha$ (see [Remark 4.2, \cite{wanduku-biomath}]), it follows from (\ref{ch1.sec4.lemma1.eq1}) that as $(\mu+d+\alpha)\rightarrow 0^{+}$ and consequently $\frac{1}{(\mu+d+\alpha)}$ is sufficiently large, then $e^{-(\mu+d+\alpha)(\rho+1)h}\rightarrow 1^{-}$. Moreover, from (\ref{ch1.sec4.lemma1.eq1}), $v_{2}\rightarrow qI^{*}_{1}$. That is, suppose malaria is still transmitted in the population $\beta>0$, so that as described in [1.] above, more susceptible individuals become infected, but the effective average lifespan in the population is high because for example, malaria is effectively treated, and healthier ways to live are encouraged so that less people die from the disease at rate $0\leq d<<1$, and also from natural causes at rate $0\leq\mu<<1$, then the total infectious class that remains over sufficiently long time $v_{2}$ will be a fraction $q>>qe^{-(\mu+d+\alpha)(\rho+1)h}$ of all those who are infected in the steady state population $I^{*}_{1}$. In other words, a larger number of infectious people remains over time when malaria is treated effectively, and better living standards are encouraged.
        \item[3.] The question of under what conditions the population ever gets extinct in time can easily be answered from [1.]~\&~[2.] above. Since as $\beta\rightarrow +\infty$, and $(\mu+d+\alpha)\rightarrow \infty$,  then $v_{1}\rightarrow 0^{+}$, and $v_{2}\rightarrow 0^{+}$, respectively, from (\ref{ch1.sec4.lemma1.eq1}). Thus,  one concludes that extinction is ever possible in time ( that is, over sufficiently large time), whenever disease transmission rate is significantly high, and the response to malaria treatment and also the response of the population to improve living standards are very poor. These observations are investigated numerically in the next section.
        %
%%%%%%%%%%%
 \end{rem}
 %%%%%%%%%%%%%%%%%%
 %%%%%%%%%%%%%%%%%%
\section{Example for Permanence of malaria}\label{ch1.sec4}
%  It should be noted that some of the numerical examples discussed in this section are utilized in various capacities elsewhere to address different sub-objectives of the current on going project.
In this study, the examples exhibited in this section are used to facilitate understanding about the influence of malaria transmission rate represented by the rate $\beta$, and the response to the standards of malaria treatment, and human living conditions which are indicated by the disease related death ($d$)  and natural ($\mu$) deathrates in the population respectively, on the permanence of the disease in the population. This objective is realized in a simplistic manner by examining the behavior of the trajectories of  the different states ($S, E, I, R$) of the system (\ref{ch1.sec0.eq3}) in the neighborhood of  the endemic equilibrium $E_{1}=(S^{*}_{1}, E^{*}_{1},I^{*}_{1}, R^{*}_{1})$ of the system.

Recall [Theorem~5.2, Wanduku\cite{wanduku-biomath}] asserts that the endemic equilibrium $E_{1}$ exists, whenever the basic reproduction number $R^{*}_{0}>1$, where $R^{*}_{0}$ is defined in (\ref{ch1.sec2.lemma2a.corrolary1.eq4}). It follows that when the conditions of [Theorem~5.2, Wanduku\cite{wanduku-biomath}] are satisfied, then the endemic equilibrium $E_{1}=(S^{*}_{1}, E^{*}_{1},I^{*}_{1}, R^{*}_{1})$ satisfies the following system
\begin{equation}\label{ch1.sec4.eq1}
\left\{
\begin{array}{lll}
&&B-\beta Se^{-\mu T_{1}}G(I)-\mu S+\alpha I e^{-\mu T_{3}}=0\\
&&\beta Se^{-\mu T_{1}}G(I)-\mu E -\beta Se^{-\mu (T_{1}+T_{2})}G(I)=0\\
&&\beta Se^{-\mu (T_{1}+T_{2})}G(I)-(\mu+d+\alpha)I=0\\
&&\alpha I-\mu R-\alpha I e^{-\mu T_{3}}=0
\end{array}
\right.
\end{equation}
%%%
 The following convenient list of parameter values in Table~\ref{ch1.sec4.table2} are used to generate and examine the trajectories of the different states of the  system (\ref{ch1.sec0.eq3}), whenever $R^{*}_{0}>1$, and the intensities of the malaria incidence rate, and the deathrates of the population in the system continuously change. The following new variables are introduced.
    (1) The special nonlinear incidence  functions $G(I)=\frac{aI}{1+I}, a\geq 0$ in \cite{gumel} is utilized to generate the numeric results. The variable $a$ is used to scale the malaria transmission rate $\beta$, and it will be called the intensity of the disease transmission rate. The total conversion rate from infectiousness in the population $\theta(\mu+d+\alpha), \theta\geq 0$ is also scaled by the parameter $\theta\geq 0$, and this parameter shall be referred to as the response intensity of malaria regulation.

  For the set of parameter values in Table~\ref{ch1.sec4.table2}, it is easily seen that, $R^{*}_{0}=80.7854>1$. Furthermore, the endemic equilibrium for the system $E_{1}$ is given as follows:- $E_{1}=(S^{*}_{1},E^{*}_{1},I^{*}_{1})=(0.2286216,0.07157075,0.9929282)$.
  %%%%
 \begin{table}[h]
  \centering
  \caption{A list of specific values chosen for the system parameters for the examples in subsection~\ref{ch1.sec4.subsec1}}\label{ch1.sec4.table2}
  \begin{tabular}{l l l}
  Disease transmission rate&$\beta$& 0.6277\\\hline
  Constant Birth rate&$B$&$ \frac{22.39}{1000}$\\\hline
  Recovery rate& $\alpha$& 0.05067\\\hline
  Disease death rate& $d$& 0.01838\\\hline
  Natural death rate& $\mu$& $0.002433696$\\\hline
  %Intensity of fluctuations& $\sigma_{i}, i=S, E, I, R, \beta$& 0.05\\\hline
  Incubation delay time in vector& $T_{1}$& 2 units \\\hline
  Incubation delay time in host& $T_{2}$& 1 unit \\\hline
  Immunity delay time& $T_{3}$& 4 units\\\hline
  \end{tabular}
\end{table}

The Euler approximation scheme is used to generate the trajectories for the different states $S(t), E(t), I(t), R(t)$ over the time interval $[0,T]$, where $T=\max(T_{1}+T_{2}, T_{3})=4$.  Furthermore, the following initial conditions are used
\begin{equation}\label{ch1.sec4.eq1}
\left\{
\begin{array}{l l}
S(t)= 10,\\
E(t)= 5,\\
I(t)= 6,\\
R(t)= 2,
\end{array}
\right.
\forall t\in [-T,0], T=\max(T_{1}+T_{2}, T_{3})=4.
\end{equation}
%%%%%%%%%%%%%%%%%
\subsection{Example 1: Effect of the intensity $a\geq 0$ of malaria transmission rate  }\label{ch1.sec4.subsec1}%random fluctuations in the system on sample population density over time
\begin{figure}[H]
\begin{center}
\includegraphics[height=6cm]{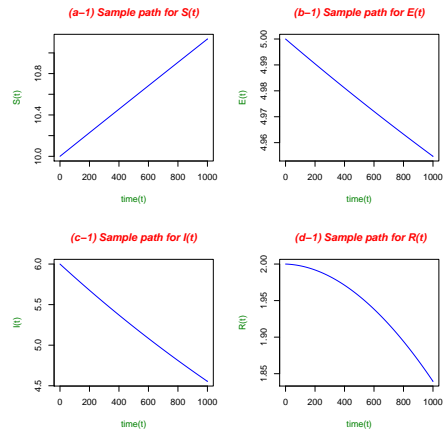}
\caption{(a-1), (b-1), (c-1) and (d-1) show the trajectories of the disease states $(S,E,I,R)$, respectively, over sufficiently long time $t=1000$, whenever the intensity of the incidence of malaria $a=0.05$, and the response intensity of malaria regulation is fixed at $\theta=1$. The broken lines represent the endemic equilibrium $E_{1}=(S^{*}_{1},E^{*}_{1},I^{*}_{1})=(5.116312,0.07157075, 1.229982)$. Furthermore, $\min{(S(t))}=10$, $\min{(E(t))}=4.954723$, $\min{(I(t))}=4.554145$ and $\min{(R(t))}=1.839505$. Moreover, the basic reproduction number in this case is $R^{*}_{0}= 4.03927>1$
}\label{ch1.sec4.subsec1.fig1}
\end{center}
\end{figure}

The effect of continuously changing the intensity $a\geq 0$ of the malaria transmission rate and consequently changing the incidence rate of malaria on the permanence of the disease is exhibited in the Figures~\ref{ch1.sec4.subsec1.fig1}-\ref{ch1.sec4.subsec1.fig2}
\begin{figure}[H]
\begin{center}
\includegraphics[height=6cm]{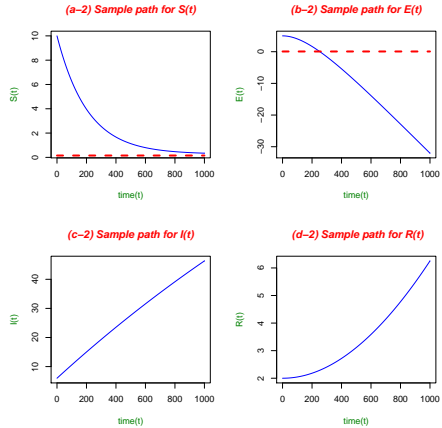}
\caption{(a-2), (b-2), (c-2) and (d-2) show the trajectories of the disease states $(S,E,I,R)$, respectively, over sufficiently long time $t=1000$, whenever the intensity of the incidence of malaria $a=1.5$, and the response intensity of malaria regulation is fixed at $\theta=1$. The broken lines represent the endemic equilibrium $E_{1}=(S^{*}_{1},E^{*}_{1},I^{*}_{1})=(0.1521307,0.07157075,0.9892183)$. Furthermore, $\min{(S(t))}=0.3449024$, $\min{(E(t))}=-32.03996$, $\min{(I(t))}=6$ and $\min{(R(t))}=1.999992$. Moreover, the basic reproduction number in this case is $R^{*}_{0}= 121.1781>1$
}\label{ch1.sec4.subsec1.fig2}
\end{center}
\end{figure}

It is easy to see from Figures~\ref{ch1.sec4.subsec1.fig1}-\ref{ch1.sec4.subsec1.fig2} that for a fixed response intensity of malaria regulation $\theta=1$, by continuously increasing the intensity of malaria transmission rate from $a=0.05$ to $a=1.5$, the smallest value of $S(t)$ denoted $\min{S(t)}$ decreases over time from $\min{S(t)}=10$ to $\min{S(t)}=0.3449024$, while the trajectory for $S(t)$ in Figures~\ref{ch1.sec4.subsec1.fig1}~$(a-1)$ changes from a monotonic increasing function to a monotonic decreasing function in Figures~\ref{ch1.sec4.subsec1.fig2}~$(a-2)$ over long time, respectively. This observation confirms the Remark~\ref{ch1.sec4.rem1}~[1.], which asserts that lower transmission rate of malaria allows a larger lower bound for the susceptible state over time and vice versa. That is, the higher malaria transmission rate which results in more susceptible people getting infected,  leads to a rise in the infectious population $I(t)$ over time, where $I((t)$ changes from a monotonic decreasing function in Figure~\ref{ch1.sec4.subsec1.fig1}~$(c-1)$ to a monotonic increasing in function in  Figure~\ref{ch1.sec4.subsec1.fig2}~$(c-2)$. Consequently, only a smaller number of susceptibles $S(t)$ remain over time.
\subsubsection{Example 2: Effect of the response intensity $\theta\geq 0$ of malaria regulation }\label{ch1.sec4.subsec1.1}
\begin{figure}[H]
\begin{center}
\includegraphics[height=6cm]{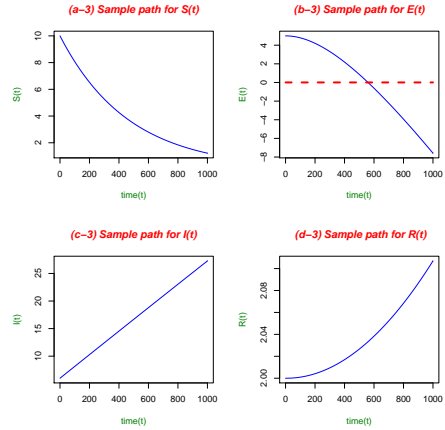}
\caption{(a-3), (b-3), (c-3) and (d-3) show the trajectories of the disease states $(S,E,I,R)$, respectively, over sufficiently long time $t=1000$, whenever the intensity of the incidence of malaria is fixed at $a=1$, and the response intensity of malaria regulation is $\theta=0.05$. The broken lines represent the endemic equilibrium $E_{1}=(S^{*}_{1},E^{*}_{1},I^{*}_{1})=(-0.03663787,0.003574402,-7.432011)$. Furthermore, $\min{(S(t))}=10$, $\min{(E(t))}=-7.586546$, $\min{(I(t))}=6$ and $\min{(R(t))}=1.999997$. Also, note that all  negative values have no meaningful interpretation, except a significance of extinction of the population if the negative values occur over sufficiently long time. Moreover, the basic reproduction number in this case is $R^{*}_{0}=32314.16>1$.
}\label{ch1.sec4.subsec1.fig3}
\end{center}
\end{figure}

\begin{figure}[H]
\begin{center}
\includegraphics[height=6cm]{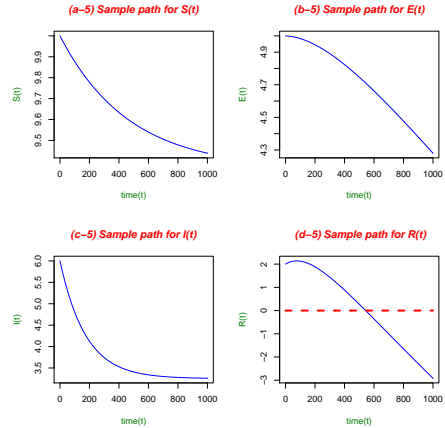}
\caption{(a-5), (b-5), (c-5) and (d-5) show the trajectories of the disease states $(S,E,I,R)$, respectively, over sufficiently long time $t=1000$, whenever the intensity of the incidence of malaria is fixed at $a=1$, and the response intensity of malaria regulation is $\theta=20$. The broken lines represent the endemic equilibrium $E_{1}=(S^{*}_{1},E^{*}_{1},I^{*}_{1})=(8.63627,1.465039,2.276611)$. Furthermore, $\min{(S(t))}=9.438989$, $\min{(E(t))}=4.280227$, $\min{(I(t))}=3.260888$ and $\min{(R(t))}=-2.913245$. Also, note that all  negative values have no meaningful interpretation, except a significance of extinction of the population if the negative values occur over sufficiently long time. Moreover, the basic reproduction number in this case is $R^{*}_{0}= 0.2019635<1$.
}\label{ch1.sec4.subsec1.fig4}
\end{center}
\end{figure}

\begin{figure}[H]
\begin{center}
\includegraphics[height=6cm]{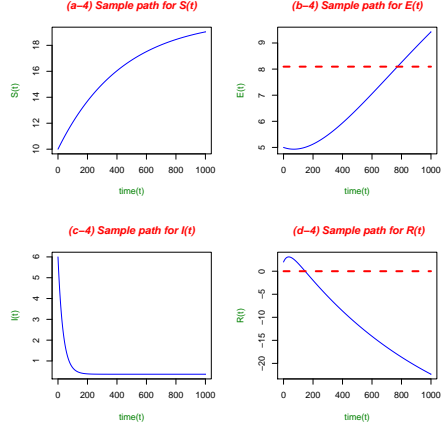}
\caption{(a-4), (b-4), (c-4) and (d-4) show the trajectories of the disease states $(S,E,I,R)$, respectively, over sufficiently long time $t=1000$, whenever the intensity of the incidence of malaria is fixed at $a=1$, and the response intensity of malaria regulation is $\theta=100$. The broken lines represent the endemic equilibrium $E_{1}=(S^{*}_{1},E^{*}_{1},I^{*}_{1})=(-67.80324,8.093293,-3.868886)$. Furthermore, $\min{(S(t))}=10$, $\min{(E(t))}=4.93237$, $\min{(I(t))}=0.3626737$ and $\min{(R(t))}=-22.35532$. Also, note that all  negative values have no meaningful interpretation, except a significance of extinction of the population if the negative values occur over sufficiently long time. Moreover, the basic reproduction number in this case is $R^{*}_{0}=0.00807854<1$.
}\label{ch1.sec4.subsec1.fig5}
\end{center}
\end{figure}

The effect of continuously changing the response intensity $\theta\geq 0$ of the malaria regulation, and consequently changing the survival probability rate of malaria patients (see Remark~\ref{ch1.sec4.rem1}[2.]) on the permanence of the disease is exhibited in the Figures~\ref{ch1.sec4.subsec1.fig3}-\ref{ch1.sec4.subsec1.fig5}
%%%%%%%
%%%%%%%
%%%%%%%%%%%%%%%%%

It is easy to see from Figures~\ref{ch1.sec4.subsec1.fig3}-\ref{ch1.sec4.subsec1.fig4} that for a fixed intensity of malaria transmission rate $a=1$, by continuously increasing the response intensity of malaria regulation from $\theta=0.05$ to $\theta=100$, the smallest value of $I(t)$ denoted $\min{I(t)}$  continuously decreases over time from $\min{I(t)}=6$ to $\min{I(t)}=0.3626737$, while the trajectory for $I(t)$ in Figures~\ref{ch1.sec4.subsec1.fig3}~$(c-3)$ changes from a monotonic increasing function to a monotonic decreasing function in Figure~\ref{ch1.sec4.subsec1.fig5}~$(c-4)$ and Figure~\ref{ch1.sec4.subsec1.fig4}~$(c-5)$ over long time, respectively. This observation confirms the Remark~\ref{ch1.sec4.rem1}~[2.], which asserts that smaller values of $\theta(\mu+d+\alpha)\rightarrow 0^{+}$, and consequently larger values for the effective average lifespan of an individual who gets infected and recovers from disease $\frac{1}{(\mu+d+\alpha)}$, allow for an eventual larger lower bound for the infectious state over time and vice versa.

Also observe that the basic reproduction number $R^{*}_{0}$ decreases across from $R^{*}_{0}=32314.16>1$ to $R^{*}_{0}=0.2019635<1$ and $R^{*}_{0}=0.00807854<1$, as  $\theta$ increases from $\theta=0.05$ to $\theta=100$, which signifies that the increase in the response to treatment or malaria control is effectively leading to eradication of the malaria parasite, which results in an increase in the number of susceptible people in the population as depicted in Figure~\ref{ch1.sec4.subsec1.fig5}~$(a-4)$, where the susceptible state $S(t)$ becomes a monotonic increasing function, changing from a monotonic decreasing function in Figure~\ref{ch1.sec4.subsec1.fig3}~$(a-3)$ and Figure~\ref{ch1.sec4.subsec1.fig4}~$(a-5)$, as $\theta$ increases in value.
\subsection{Example 3: Joint effects the intensities of malaria transmission and response to malaria regulation }
\begin{figure}[H]
\begin{center}
\includegraphics[height=6cm]{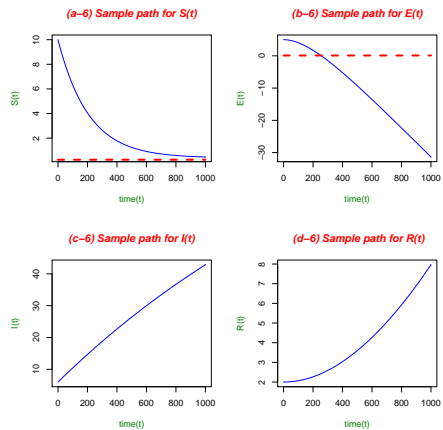}
\caption{(a-6), (b-6), (c-6) and (d-6) show the trajectories of the disease states $(S,E,I,R)$, respectively, over sufficiently long time $t=1000$, whenever the intensity of the incidence of malaria is $a=1.5$, and the response intensity of malaria regulation is $\theta=1.5$. The broken lines represent the endemic equilibrium $E_{1}=(S^{*}_{1},E^{*}_{1},I^{*}_{1})=(0.247531,0.1074215,1.149901)$. Furthermore, $\min{(S(t))}=0.467897$, $\min{(E(t))}=-31.35931$, $\min{(I(t))}=6$ and $\min{(R(t))}=1.999997$. Also, note that all  negative values have no meaningful interpretation, except a significance of extinction of the population if the negative values occur over sufficiently long time. Moreover, the basic reproduction number in this case is $R^{*}_{0}= 53.85694>1$.
}\label{ch1.sec4.subsec1.fig6}
\end{center}
\end{figure}

\begin{figure}[H]
\begin{center}
\includegraphics[height=6cm]{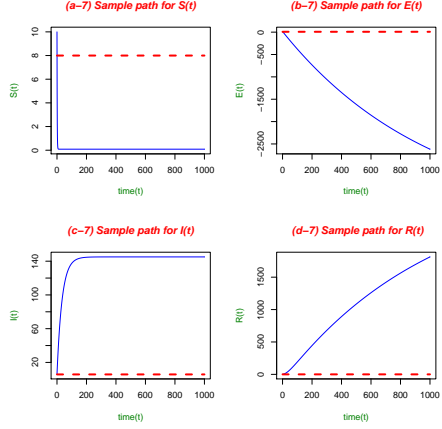}
\caption{(a-7), (b-7), (c-7) and (d-7) show the trajectories of the disease states $(S,E,I,R)$, respectively, over sufficiently long time $t=1000$, whenever the intensity of the incidence of malaria is $a=20$, and the response intensity of malaria regulation is $\theta=100$. The broken lines represent the endemic equilibrium $E_{1}=(S^{*}_{1},E^{*}_{1},I^{*}_{1})=(7.998663,8.093293,5.768778)$. Furthermore, $\min{(S(t))}= 0.08683528$, $\min{(E(t))}=-2617.125$, $\min{(I(t))}=6$ and $\min{(R(t))}=2$. Also, note that all  negative values have no meaningful interpretation, except a significance of extinction of the population if the negative values occur over sufficiently long time. Moreover, the basic reproduction number in this case is $R^{*}_{0}= 0.1615708<1$.
}\label{ch1.sec4.subsec1.fig7}
\end{center}
\end{figure}

The joint effects of continuously changing the response intensity $\theta\geq 0$ of the malaria regulation, and consequently changing the survival probability rate of malaria patients, as well as continuously changing the intensity $a\geq 0$ of malaria transmission rate,  on the permanence of the disease is exhibited in the Figures~\ref{ch1.sec4.subsec1.fig6}-\ref{ch1.sec4.subsec1.fig7}
%%%%%%%%%%%%%%%%%
%%%%%%%%%%%%%%%%%
%%%%%
Observe from Figures~\ref{ch1.sec4.subsec1.fig6}-\ref{ch1.sec4.subsec1.fig7} that as the values of $\theta\geq 0$ and $a\geq 0$ continuously increase from $(\theta, a)=(1.5,1.5)$ to $(\theta, a)=(100,20)$, the trajectory for $S(t)$, continuously decreases monotonically over time with a continuously decreasing minimum value  from $\min{(S(t))}=0.467897$ to $\min{(S(t))}=0.08683528$, while the trajectory for $I(t)$ continuously increases monotonically over time and saturates at a maximum  value $\max{(I(t))}= 145.0695$ exhibited in Figure~\ref{ch1.sec4.subsec1.fig7}~$(c-7)$. In addition, the trajectory for the removed class $R(t)$, monotonically increases over time with minimum and maximum values rising in the ranges $(\min{(R(t))},\max{(R(t))})=(2,7.964306)$ and   $(\min{(R(t))},\max{(R(t))})=(2,1812.767)$ in Figures~\ref{ch1.sec4.subsec1.fig6}-\ref{ch1.sec4.subsec1.fig7}, respectively. Furthermore, the basic reproduction number also decreases from $R^{*}_{0}= 53.85694>1$ to $R^{*}_{0}= 0.1615708<1$, respectively.

 These observations suggests that increasing the intensity of the malaria transmission rate, and also increasing the response intensity to malaria regulation results in more people removed from the disease with natural immunity than they are infected,  if the intensity of the response to malaria control $\theta$ is higher than the intensity of malaria transmission $a$. Indeed, this is evident from Figure~\ref{ch1.sec4.subsec1.fig6}$(d-6)$ and Figure~\ref{ch1.sec4.subsec1.fig7}$(d-7)$, where $R(t)$ continuously increases as $(\theta, a)$ continuously increases from $(\theta, a)=(1.5,1.5)$ to $(\theta, a)=(100,20)$. The rise in the infectious population $I(t)$ in Figure~\ref{ch1.sec4.subsec1.fig6}$(c-6)$ and Figure~\ref{ch1.sec4.subsec1.fig7}$(c-7)$, which saturates in Figure~\ref{ch1.sec4.subsec1.fig7}$(c-7)$, is attributed to the increase in disease transmission rate as $a\geq 0$ increases from $a=1.5$ to $a=20$. However, since $\theta=100>> 20=a$, signifying that the response to malaria regulation is significantly larger than the disease transmission rate, it implies that recovery from disease with natural immunity $R(t)$ over time dominates the transmission of malaria to susceptible people $S(t)$ over time. Finally, despite the rise in malaria transmission as $a\geq 0$ increases from $a=1.5$ to $a=20$,  the decline in the basic reproduction number $R^{*}_{0}$ from $R^{*}_{0}= 53.85694>1$ to $R^{*}_{0}= 0.1615708<1$, signifies that the malaria parasite is getting eradicated as more people respond positively to malaria treatment, and are removed with natural immunity.
\section{Extinction  of disease}\label{ch1.sec2b}
 In this section, the extinction of malaria from the system (\ref{ch1.sec0.eq3}) is investigated. It will be shown that the extinction of the  disease from the population depends only on the  basic reproduction number $R^{*}_{0}$ in (\ref{ch1.sec2.lemma2a.corrolary1.eq4}) and (\ref{ch1.sec2.theorem1.corollary1.eq3}), or on the expected survival probability rate $E(e^{-(\mu_{v}T_{1}+\mu T_{2})})$ of the malaria parasites, where $\mu_{v}$ is the natural death rate of the mosquitoes, over the complete life cycle of the parasites of length $T_{1}+T_{2}$.
%%%%
The following lemmas will be used to establish the extinction results, whenever the conditions of [Theorem~3.1, Wanduku\cite{wanduku-biomath}] hold.
\begin{lemma}\label{ch1.sec2b.lemma1}
Suppose the conditions of [Theorem~3.1, Wanduku\cite{wanduku-biomath}] hold, then the unique solution $Y(t)\in D(\infty),t\geq t_{0}$ of the  system (\ref{ch1.sec0.eq3}) also lies in the space
\begin{equation}\label{ch1.sec2b.lemma1.eq1}
  D^{expl}(\infty)=\left\{Y(t)\in \mathbb{R}^{4}_{+}:\frac{B}{\mu+d}\leq N(t)=S(t)+ E(t)+ I(t)+ R(t)\leq \frac{B}{\mu}, \forall t\in (-\infty, \infty) \right\},
  %=\bar{B}^{(-\infty, \tau_{e}]}_{\mathbb{R}^{4}_{+},}\left(0,\frac{B}{\mu}\right),
\end{equation}
where $D^{expl}(\infty)\subset D(\infty)$.  Moreover, the space $D^{expl}(\infty)$ is also self-invariant with respect to the system (\ref{ch1.sec0.eq3}).
\end{lemma}
Proof:\\
Suppose [Theorem~3.1, Wanduku\cite{wanduku-biomath}] holds, then it follows from (\ref{ch1.sec0.eq3}) and (\ref{ch1.sec0.eq13b}) that the total population $N(t)=S(t)+E(t)+I(t)+R(t)$ satisfies the following inequality
\begin{equation}\label{ch1.sec2b.lemma1.proof.eq1}
[B-(\mu+d)N(t)]dt\leq dN(t)\leq [B-(\mu)N(t)]dt.
\end{equation}
It is easy to see from (\ref{ch1.sec2b.lemma1.proof.eq1}) that
\begin{equation}\label{ch1.sec2b.lemma1.proof.eq2}
\frac{B}{\mu+d}\leq \liminf_{t\rightarrow \infty}N(t)\leq \limsup_{t\rightarrow \infty}N(t)\leq \frac{B}{\mu},
\end{equation}
and (\ref{ch1.sec2b.lemma1.eq1}) follows immediately.
\begin{lemma}\label{ch1.sec2b.lemma2}
Let [Theorem~3.1, Wanduku\cite{wanduku-biomath}] hold, and define the following Lyapunov functional in $D^{expl}(\infty)$,
\begin{eqnarray}
\tilde{V}(t)&=&V(t)+\beta\left[\int_{t_{0}}^{h_{2}}\int_{t_{0}}^{h_{1}}f_{T_{2}}(u)f_{T_{1}}(s)e^{-(\mu_{v}s+\mu u)}\int^{t}_{t-u}S(\theta)\frac{G(I(\theta-s))}{I(t)}d\theta dsdu\right.\nonumber\\
&&\left. +\int_{t_{0}}^{h_{2}}\int_{t_{0}}^{h_{1}}f_{T_{2}}(u)f_{T_{1}}(s)e^{-(\mu_{v}s+\mu u)}\int^{t}_{t-s}S(t)\frac{G(I(\theta))}{I(t)}d\theta\right],
\label{ch1.sec2b.lemma2.eq1}
\end{eqnarray}
where $V(t)=\log{I(t)}$. It follows that
\begin{equation}\label{ch1.sec2b.lemma2.eq2}
\limsup_{t\rightarrow \infty}\frac{1}{t}\log{(I(t))}\leq \beta \frac{B}{\mu}E(e^{-(\mu_{v}T_{1}+\mu T_{2})})-(\mu+d+\alpha).
\end{equation}
\end{lemma}
Proof:\\
The differential operator $\dot{V}$ applied to the Lyapunov functional $\tilde{V}(t)$
with respect to the system (\ref{ch1.sec0.eq3}) leads to the following
\begin{equation}\label{ch1.sec2b.lemma2.proof.eq1}
  \dot{\tilde{V}}(t)=\beta\int_{t_{0}}^{h_{2}}f_{T_{2}}(u)\int^{h_{1}}_{t_{0}}f_{T_{1}}(s) e^{-(\mu_{v} s+\mu u)}S(t)\frac{G(I(t))}{I(t)}dsdu-(\mu+ d+ \alpha)%f(S, I)dt  + \sigma_{\beta} \int_{t_{0}}^{h_{2}}\int^{h_{1}}_{t_{0}}f_{T_{2}}(u)f_{T_{1}}(s) e^{-(\mu_{v} s+\mu u)}S(t-u)\frac{G(I(t-s-u))}{I(t)}dsdudw_{\beta}(t),
\end{equation}
%where,
%\begin{eqnarray}
%  f(S, I)&=&\beta\int_{t_{0}}^{h_{2}}f_{T_{2}}(u)\int^{h_{1}}_{t_{0}}f_{T_{1}}(s) e^{-(\mu_{v} s+\mu u)}S(t)\frac{G(I(t))}{I(t)}dsdu-(\mu+ d+ \alpha)\nonumber\\
%  &&-\frac{1}{2}\sigma^{2}_{\beta}\left(\int_{t_{0}}^{h_{2}}f_{T_{2}}(u)\int^{h_{1}}_{t_{0}}f_{T_{1}}(s) e^{-(\mu_{v} s+\mu u)}S(t-u)\frac{G(I(t-s-u))}{I(t)}dsdu\right)^{2}.\label{ch1.sec2b.lemma2.proof.eq2}
%\end{eqnarray}
Since $S(t), I(t)\in D^{expl}(\infty)$, and $G$ satisfies the conditions of Assumption~\ref{ch1.sec0.assum1}, it  follows easily from (\ref{ch1.sec2b.lemma2.proof.eq1}) that
\begin{equation}\label{ch1.sec2b.lemma2.proof.eq3}
\dot{\tilde{V}}(t)\leq \beta \frac{B}{\mu}E(e^{-(\mu_{v}T_{1}+\mu T_{2})})-(\mu+d+\alpha).
\end{equation}
Now, integrating both sides of  (\ref{ch1.sec2b.lemma2.proof.eq3}) over the interval $[t_{0},t]$, it follows from (\ref{ch1.sec2b.lemma2.proof.eq3})  and (\ref{ch1.sec2b.lemma2.eq1}) that
%%%%%%%%%%%%%%%%%%
 \begin{eqnarray}
   \log{I(t)}&\leq&\tilde{V}(t)\nonumber\\
    &\leq&\tilde{V}(t_{0})+\left[\beta \frac{B}{\mu}E(e^{-(\mu_{v}T_{1}+\mu T_{2})})-(\mu+d+\alpha)\right](t-t_{0}).\label{ch1.sec2b.lemma2.proof.eq4}%+M_{2}(t),
 \end{eqnarray}
%where $M_{2}(t)$ is defined in (\ref{ch1.sec2a.thm1.proof.eq5}).
 Diving both sides of (\ref{ch1.sec2b.lemma2.proof.eq4}) by $t$, and  taking the limit supremum  as $t\rightarrow\infty$, it is easy to see that (\ref{ch1.sec2b.lemma2.proof.eq4}) reduces to
 \begin{eqnarray}
   \limsup_{t\rightarrow\infty}\frac{1}{t}\log{I(t)}&\leq& \left[\beta \frac{B}{\mu}E(e^{-(\mu_{v}T_{1}+\mu T_{2})})-(\mu+d+\alpha)\right].\label{ch1.sec2b.lemma2.proof.eq5}
   %+\limsup_{t\rightarrow\infty}\frac{1}{t}M_{2}(t).
 \end{eqnarray}
 And the result (\ref{ch1.sec2b.lemma2.eq2}) follows immediately from  (\ref{ch1.sec2b.lemma2.proof.eq5}).

The conditions for extinction of the infectious population over time can be expressed in terms of two important parameters for the disease dynamics namely - (1) the basic reproduction number $R^{*}_{0}$ in (\ref{ch1.sec2.lemma2a.corrolary1.eq4}), and (2) the expected survival probability rate of the parasites $E(e^{-(\mu_{v}T_{1}+\mu T_{2})})$, defined in [Theorem~5.1, Wanduku\cite{wanduku-biomath}].
\begin{thm}\label{ch1.sec2b.thm1}
Suppose the conditions for Lemma~\ref{ch1.sec2b.lemma2} are satisfied, and let the basic reproduction number $R^{*}_{0}$ be defined as in (\ref{ch1.sec2.lemma2a.corrolary1.eq4}). In addition, let one of the following conditions hold
\item[1.] $R^{*}_{0}\geq 1$ and $E(e^{-(\mu_{v}T_{1}+\mu T_{2})})<\frac{1}{R^{*}_{0}}$, or
\item[2.]$R^{*}_{0}<1$.\\
Then
\begin{equation}\label{ch1.sec2b.thm1.eq1}
\limsup_{t\rightarrow \infty}\frac{1}{t}\log{(I(t))}<-\lambda.
\end{equation}
where  $\lambda>0$ is some positive constant. In other words, $I(t)$ converges to zero exponentially.
\end{thm}
Proof:\\
Suppose Theorem~\ref{ch1.sec2b.thm1}~[1.] holds, then from (\ref{ch1.sec2b.lemma2.eq2}),
\begin{equation}\label{ch1.sec2b.thm1.eq1.proof.eq1}
\limsup_{t\rightarrow \infty}\frac{1}{t}\log{(I(t))}< \beta\frac{B}{\mu}\left(E(e^{-(\mu_{v}T_{1}+\mu T_{2})})- \frac{1}{R^{*}_{0}} \right)\equiv -\lambda,
\end{equation}
where the positive constant $\lambda>0$ is taken to be  as follows
\begin{equation}\label{ch1.sec2b.thm1.eq1.proof.eq1.eq1}
\lambda\equiv(\mu+d+\alpha)-\beta \frac{B}{\mu}E(e^{-(\mu_{v}T_{1}+\mu T_{2})})=\beta\frac{B}{\mu}\left( \frac{1}{R^{*}_{0}}-E(e^{-(\mu_{v}T_{1}+\mu T_{2})}) \right)>0.
\end{equation}
 Also, suppose Theorem~\ref{ch1.sec2b.thm1}~[2.] holds, then from (\ref{ch1.sec2b.lemma2.eq2}),
\begin{eqnarray}
\limsup_{t\rightarrow \infty}\frac{1}{t}\log{(I(t))}&\leq& \beta \frac{B}{\mu}E(e^{-(\mu_{v}T_{1}+\mu T_{2})})-(\mu+d+\alpha)\nonumber\\
&<& \beta \frac{B}{\mu}-(\mu+d+\alpha)= -(1-R^{*}_{0})(\mu+d+\alpha)\equiv -\lambda,\label{ch1.sec2b.thm1.eq1.proof.eq2}
\end{eqnarray}
where the positive constant $\lambda>0$ is taken to be  as follows
\begin{equation}\label{ch1.sec2b.thm1.eq1.proof.eq2.eq1}
\lambda\equiv(1-R^{*}_{0})(\mu+d+\alpha)>0.
\end{equation}
%%%
\begin{rem}
Theorem~\ref{ch1.sec2b.thm1}, [Theorem~3.1, Wanduku\cite{wanduku-biomath}] and Lemma~\ref{ch1.sec2b.lemma1} signify that all trajectories of the solution $(S(t), I(t))$  of the decoupled system containing equations for  $S$ and $I$ in the system  (\ref{ch1.sec0.eq3}) that start  in $D^{expl}(\infty)\subset D(\infty)$ remain bounded in $D^{expl}(\infty)$. Moreover, on the phase plane of the solution $(S(t), I(t))$, the trajectory of the  infectious state $I(t), t\geq t_{0}$ ultimately turn to zero exponentially, whenever either the expected survival probability rate of the malaria parasites satisfy  $E(e^{-(\mu_{v}T_{1}+\mu T_{2})})<\frac{1}{R^{*}_{0}}$ for $R^{*}_{0}\geq 1$ , or whenever the basic production number of the disease satisfy $R^{*}_{0}<1$.  Furthermore, the  Lyapunov exponent from (\ref{ch1.sec2b.thm1.eq1}) is estimated by the  term $\lambda$, defined in (\ref{ch1.sec2b.thm1.eq1.proof.eq1.eq1}) and (\ref{ch1.sec2b.thm1.eq1.proof.eq2.eq1}).

%%
  %The condition (\ref{ch1.sec2a.thm1.eq1}) for extinction of the disease relates the intensities of the random fluctuations in the disease transmission and the infectious population natural death rates $\sigma^{2}_{\beta}$ and $\sigma^{2}_{I}$, respectively. Let $Q(\sigma^{2}_{\beta},\sigma^{2}_{I})=\frac{\beta^{2}}{2\sigma^{2}_{\beta}}-(\mu + d+ \alpha + \frac{1}{2}\sigma^{2}_{I})$ be called the extinction function, then
  %%%%%
  It follows from (\ref{ch1.sec2b.thm1.eq1}) that when either of the conditions in Theorem~\ref{ch1.sec2b.thm1}[1.-2.]  hold, then the infectious population $I(t)$ dies out exponentially, whenever  $\lambda$ in (\ref{ch1.sec2b.thm1.eq1.proof.eq1.eq1}) and (\ref{ch1.sec2b.thm1.eq1.proof.eq2.eq1}) is positive, that is, $\lambda>0$. In addition, the rate of the exponential decay of each trajectories of the infectious population $I(t)$ in each scenario of Theorem~\ref{ch1.sec2b.thm1}[1.-2.] is given by the  estimate $\lambda>0$ of the Lyapunov exponent in (\ref{ch1.sec2b.thm1.eq1.proof.eq1.eq1}) and (\ref{ch1.sec2b.thm1.eq1.proof.eq2.eq1}).

  The conditions in  Theorem~\ref{ch1.sec2b.thm1}[1.-2.] can also be interpreted as follows. Recall [\cite{wanduku-biomath}, Remark~4.2], the basic reproduction number   $R^{*}_{0}$ in (\ref{ch1.sec2.lemma2a.corrolary1.eq4}) (similarly in (\ref{ch1.sec2.theorem1.corollary1.eq3})) represents the expected number of secondary malaria cases that result from one infective placed in the steady state disease free population $S^{*}_{0}=\frac{B}{\mu}$. Thus, $\frac{1}{R^{*}_{0}}=\frac{(\mu+d+\alpha)}{\beta S^{*}_{0}}$, for $R^{*}_{0}\geq 1$, represents the probability rate of  infectious persons in the secondary infectious population $\beta S^{*}_{0}$ leaving the infectious state either through natural death $\mu$, diseases related death $d$, or recovery and acquiring natural immunity at the rate $\alpha$. Thus, $\frac{1}{R^{*}_{0}}$ is the effective probability rate of surviving infectiousness until recovery with acquisition of natural immunity. Moreover, $\frac{1}{R^{*}_{0}}$ is a probability measure provided $R^{*}_{0}\geq 1$.

  In addition, recall [\cite{wanduku-biomath}, Theorem~5.1\&5.2] asserts that when $R^{*}_{0}\geq 1$, and the expected survival probability $E(e^{-(\mu_{v}T_{1}+\mu T_{2})})$ is significantly large, then the outbreak of  malaria establishes a malaria endemic steady state population $E_{1}$. The conditions for extinction of disease in  Theorem~\ref{ch1.sec2b.thm1}[1.], that is $R^{*}_{0}\geq 1$ and $E(e^{-(\mu_{v}T_{1}+\mu T_{2})})<\frac{1}{R^{*}_{0}}$ suggest that in the event where $R^{*}_{0}\geq 1$, and the disease is aggressive, and likely to establish an endemic steady state population, if the expected survival probability rate $E(e^{-(\mu_{v}T_{1}+\mu T_{2})})$ of the malaria parasites over their complete life cycle of length $T_{1}+T_{2}$, is less than $\frac{1}{R^{*}_{0}}$- the effective probability rate of surviving infectiousness until recovery with natural immunity, then the malaria epidemic fails to establish an endemic steady state, and as a result, the disease ultimately dies out at an exponential rate $\lambda$ in (\ref{ch1.sec2b.thm1.eq1.proof.eq1.eq1}).
  %%%
  This result suggests that, malaria control policies should focus on vector control strategies such as genetic modification techniques in order to reduce the chances of survival of the malaria parasites inside the mosquitos, and in the human beings.

   In the event where  $R^{*}_{0}< 1$ in  Theorem~\ref{ch1.sec2b.thm1}[2.], extinction of disease occurs exponentially over sufficiently long time, regardless of the survival of the parasites. Moreover, the rate of extinction is  $\lambda$ in (\ref{ch1.sec2b.thm1.eq1.proof.eq2.eq1}).
%%
 %Also observe that the conditions in Theorem~\ref{ch1.sec2b.thm1}[1.-2.] for extinction of the infectious population $I(t)$ in the case of noise originating exclusively from the disease transmission rate $\beta$ has no bearings on the intensity of the  noise from the disease transmission rate $\sigma_{\beta}$.
%   %The function $Q$ can be used to evaluate the qualitative effects of the intensities $\sigma_{\beta}$ and $\sigma_{I}$ on the extinction of the disease from the system. Indeed, observe that the function $Q$ increases monotonically with respect to continuous changes in each intensity $\sigma_{\beta}$ and $\sigma_{I}$. This observation suggests that larger values of the intensities $\sigma_{\beta}$ and $\sigma_{I}$, lead to large values of $Q$, and consequently, a larger rate of  extinction of the disease from the population. Figure~\ref{ch1.sec2a.rem1.figure1} illustrates the behavior of the decay rate $Q$, as the intensities  $\sigma_{\beta}$ and $\sigma_{I}$ of the independent white noise processes in the system continuously increase in value.
\end{rem}
Theorem~\ref{ch1.sec2b.thm1} characterizes the behavior of the trajectories of the $I(t)$ coordinate of the solution $(S(t),I(t))$ of the decoupled system containing equations for  $S$ and $I$ in the system  (\ref{ch1.sec0.eq3}) in the phase plane. The question remains about how the trajectories for the $S(t)$ behave asymptotically in the phase plane.
   %
   %As it can be observed  from  several simulation studies involving white noise processes, in many occasions, the extinction of the infectious population over time coincides with extinction of the susceptible population, if the intensity of the noise in the epidemic dynamic system is high. And this suggests that the extinction of the disease from the population does not always imply the survival of the disease free population over time.

 The following result describes the average behavior of the trajectories of the susceptible population  $S(t)$ over sufficiently long time, and also states conditions for the asymptotic stability in the mean, in the event where the conditions of Theorem~\ref{ch1.sec2b.thm1} are satisfied.
\begin{thm}\label{ch1.sec2b.thm2}
Suppose any of the conditions in the hypothesis of Theorem~\ref{ch1.sec2b.thm1}[1.-2.] are satisfied. It follows that in $D^{expl}(\infty)$, the trajectories of the susceptible state $S(t)$ of the solution $(S(t), I(t))$  of the decoupled system containing equations for  $S$ and $I$ in the system  (\ref{ch1.sec0.eq3})  satisfy
\begin{equation}\label{ch1.sec2b.thm2.eq1}
\lim_{t\rightarrow \infty}\frac{1}{t}\int_{t_{0}}^{t}S(\xi)d\xi=\frac{B}{\mu}.
\end{equation}
That is, the susceptible population is persistent over long-time in the mean (see definition of  persistence in the mean in \cite{zhien}), and hence, asymptotically stable. Moreover, the average value of the susceptible population over sufficiently long time is the disease-free equilibrium $S^{*}_{0}=\frac{B}{\mu}$.
\end{thm}
Proof:\\
Suppose either of the conditions in Theorem~\ref{ch1.sec2b.thm1}[1.-2.] hold, then  it follows clearly from Theorem~\ref{ch1.sec2b.thm1} that  for every $\epsilon>0$, there is a positive constant $K_{1}(\epsilon)\equiv K_{1}>0$, such that
\begin{equation}\label{ch1.sec2b.thm2.proof.eq2}
  I(t)<\epsilon,\quad  \textrm{whenever $t>K_{1}$}.
\end{equation}
It follows from (\ref{ch1.sec2b.thm2.proof.eq2}) that
\begin{equation}\label{ch1.sec2b.thm2.proof.eq3}
  I(t-s)<\epsilon,\quad \textrm{whenever $t>K_{1}+h_{1},\forall s\in [t_{0}, h_{1}]$}.
\end{equation}
In $D^{expl}(\infty)$, define
\begin{equation}\label{ch1.sec2b.thm2.proof.eq4}
V_{1}(t)=S(t)+\alpha \int_{t_{0}}^{\infty} f_{T_{3}}(r)e^{\mu r}\int_{t-r}^{t}I(\theta)d\theta dr.
\end{equation}
The differential operator $\dot{V}_{1}$ applied to the Lyapunov functional $V_{1}(t)$ in (\ref{ch1.sec2b.thm2.proof.eq4}) leads to the following
\begin{equation}\label{ch1.sec2b.thm2.proof.eq5}
  \dot{V}_{1}(t)=g(S, I)-\mu S(t),%dt-\sigma_{\beta} S(t)\int^{h_{1}}_{t_{0}}f_{T_{1}}(s) e^{-\mu_{v} s}G(I(t-s))dsdw_{\beta}(t),
\end{equation}
where
\begin{equation}\label{ch1.sec2b.thm2.proof.eq6}
  g(S,I)=B-\beta S(t)\int^{h_{1}}_{t_{0}}f_{T_{1}}(s) e^{-\mu_{v} s}G(I(t-s))ds+\alpha E(e^{-\mu T_{3}}) I(t).
\end{equation}
Estimating the right-hand-side of (\ref{ch1.sec2b.thm2.proof.eq5}) in $D^{expl}(\infty)$, and integrating over $[t_{0},t]$, it follows from (\ref{ch1.sec2b.thm2.proof.eq2})-(\ref{ch1.sec2b.thm2.proof.eq3}) that
\begin{eqnarray}
  V_{1}(t)&\leq& V_{1}(t_{0})+B(t-t_{0})+\int_{t_{0}}^{K_{1}}\alpha I(\xi)d\xi +\int_{K_{1}}^{t}\alpha I(\xi)d\xi -\mu\int_{t_{0}}^{t} S(\xi)d\xi,\nonumber\\
  &\leq& V_{1}(t_{0})+B(t-t_{0})+\alpha \frac{B}{\mu}(K_{1}-t_{0})  +\alpha (t-K_{1})\epsilon -\mu\int_{t_{0}}^{t} S(\xi)d\xi.\label{ch1.sec2b.thm2.proof.eq7}
\end{eqnarray}
%where
%\begin{equation}\label{ch1.sec2b.thm2.proof.eq8}
%  M_{3}(t)=\sigma_{\beta}\int_{t_{0}}^{t} S(\xi)\int^{h_{1}}_{t_{0}}f_{T_{1}}(s) e^{-\mu_{v} s}G(I(\xi-s))dsdw_{\beta}(\xi).
%\end{equation}
%Observe that similarly to (\ref{ch1.sec2b.lemma2.proof.eq5a})-(\ref{ch1.sec2b.lemma2.proof.eq8}), it is easy to see by the strong law of large numbers for local martingales (see, e.g. \cite{mao}) that
%\begin{equation}\label{ch1.sec2b.thm2.proof.eq9}
%  \lim_{t\rightarrow \infty}{\frac{1}{t}M_{3}(t)}=0,\quad a.s.
%\end{equation}
Thus, dividing both sides of (\ref{ch1.sec2b.thm2.proof.eq7}) by $t$ and taking the limit supremum as $t\rightarrow \infty$, it follows that
\begin{equation}\label{ch1.sec2b.thm2.proof.eq10}
\limsup_{t\rightarrow \infty}\frac{1}{t}\int_{t_{0}}^{t}S(\xi)d\xi\leq \frac{B}{\mu}+ \frac{\alpha}{\mu}\epsilon.%\quad a.s.
\end{equation}
On the other hand, estimating $g(S,I)$ in (\ref{ch1.sec2b.thm2.proof.eq6}) from below and using the conditions of Assumption~\ref{ch1.sec0.assum1} and (\ref{ch1.sec2b.thm2.proof.eq3}), it is easy to see that  in $D^{expl}(\infty)$,
\begin{eqnarray}
  g(S,I) &\geq & B-\beta S(t)\int^{h_{1}}_{t_{0}}f_{T_{1}}(s) e^{-\mu_{v} s}(I(t-s))ds\nonumber \\
   &\geq& B-\beta \frac{B}{\mu}E(e^{-\mu_{v}T_{1}})\epsilon,\forall t>K_{1}+h_{1},\nonumber\\
   &\geq&B-\beta \frac{B}{\mu}\epsilon.\label{ch1.sec2b.thm2.proof.eq11}
\end{eqnarray}
Moreover, for $t\in[t_{0}, K_{1}+h_{1}]$, then
\begin{equation}\label{ch1.sec2b.thm2.proof.eq11.eq1}
  g(S,I)\geq B-\beta \left(\frac{B}{\mu}\right)^{2}.
\end{equation}
Therefore, applying (\ref{ch1.sec2b.thm2.proof.eq11})-(\ref{ch1.sec2b.thm2.proof.eq11.eq1}) into (\ref{ch1.sec2b.thm2.proof.eq5}), then integrating both sides of (\ref{ch1.sec2b.thm2.proof.eq5}) over $[t_{0},t]$, and diving the result by $t$,  it is easy to see from (\ref{ch1.sec2b.thm2.proof.eq5}) that
\begin{equation}\label{ch1.sec2b.thm2.proof.eq12}
  \frac{1}{t}V_{1}(t)\geq  \frac{1}{t}V_{1}(t_{0})+B(1-\frac{t_{0}}{t}) -\frac{1}{t}\beta\left( \frac{B}{\mu}\right)^{2}(K_{1}+h_{1}-t_{0})-\beta\frac{B}{\mu}\epsilon[1-\frac{K_{1}+h_{1}}{t}] -\frac{1}{t}\mu\int_{t_{0}}^{t}S(\xi)d\xi.%-\frac{1}{t}M_{3}(t).
\end{equation}
Observe that in  $D^{expl}(\infty)$, $\lim_{t\rightarrow \infty}\frac{1}{t}V_{1}(t)=0$, and $\lim_{t\rightarrow \infty}\frac{1}{t}V_{1}(t_{0})=0$.  Therefore, rearranging (\ref{ch1.sec2b.thm2.proof.eq12}), and taking the limit infinimum of both sides as $t\rightarrow \infty$, it is easy to see that
\begin{equation}\label{ch1.sec2b.thm2.proof.eq13}
 \liminf_{t\rightarrow \infty} \frac{1}{t}\int_{t_{0}}^{t}S(\xi)d\xi\geq   \frac{B}{\mu}-\frac{1}{\mu}\beta \frac{B}{\mu}\epsilon.
\end{equation}
It follows from (\ref{ch1.sec2b.thm2.proof.eq10}) and (\ref{ch1.sec2b.thm2.proof.eq13}) that
\begin{equation}\label{ch1.sec2b.thm2.proof.eq14}
\frac{B}{\mu}-\frac{1}{\mu}\beta \frac{B}{\mu}\epsilon\leq \liminf_{t\rightarrow \infty} \frac{1}{t}\int_{t_{0}}^{t}S(\xi)d\xi\leq \limsup_{t\rightarrow \infty}\frac{1}{t}\int_{t_{0}}^{t}S(\xi)d\xi\leq \frac{B}{\mu}+ \frac{\alpha}{\mu}\epsilon.
\end{equation}
Hence, for $\epsilon$ arbitrarily small,  the result in (\ref{ch1.sec2b.thm2.eq1}) follows immediately from (\ref{ch1.sec2b.thm2.proof.eq14}). Note, the asymptotic stability of the disease-free equilibrium state for $S(t)=S^{*}_{0}$ is explained further in the Theorem~\ref{ch1.sec2b.thm3}.%Remark~\ref{ch1.sec2b.thm2.rem1}.
%%%%,
\begin{thm}\label{ch1.sec2b.thm3}
Suppose any of the conditions in the hypothesis of Theorem~\ref{ch1.sec2b.thm1}[1.-2.] are satisfied. Also, suppose the conditions of Theorem~\ref{ch1.sec2b.thm2} hold. It follows that in $D^{expl}(\infty)$, the disease-free equilibrium  for the susceptible state $S(t)$, denoted $S^{*}_{0}=\frac{B}{\mu}$,  and the infectious state $I(t)$, denoted as $I^{*}_{0}=0$,  of the solution $(S(t), I(t))$  of the decoupled system containing equations for  $S$ and $I$ in the system  (\ref{ch1.sec0.eq3})  is asymptotically stable. That is,  $E_{0}=(S^{*}_{0}, I^{*}_{0})=(\frac{B}{\mu},0)$ is asymptotically stable.
%That is, the susceptible population is persistent over long-time in the mean (see definition of  persistence in the mean in \cite{zhien}), and hence, asymptotically stable. Moreover, the average value of the susceptible population over sufficiently long time is the disease-free equilibrium $S^{*}_{0}=\frac{B}{\mu}$.
\end{thm}
%%%
Proof:\\
From [Theorem~4.1,\cite{wanduku-biomath}], $E_{0}=(S^{*}_{0}, I^{*}_{0})=(\frac{B}{\mu},0)$ is clearly a disease-free equilibrium. It is left to show that every trajectory that starts near $E_{0}$ remains near $E_{0}$, and converges asymptotically at $E_{0}$. Indeed, clearly if the hypothesis of Theorem~\ref{ch1.sec2b.thm1}[1.-2.] hold, then all trajectories in the phase-plane for the infectious state $I(t)$ converge asymptotically and exponentially to $I^{*}_{0}=0$. It is left to show that if the trajectories of the susceptible state $S(t)$ from Theorem~\ref{ch1.sec2b.thm2} (\ref{ch1.sec2b.thm2.eq1}), converge asymptotically in the mean to $S^{*}_{0}=\frac{B}{\mu}$, then they converge asymptotically to $S^{*}_{0}=\frac{B}{\mu}$.

Indeed, if on the contrary, there exist a trajectory for $S(t)$ starting near $S^{*}_{0}=\frac{B}{\mu}$ that does not stay near $S^{*}_{0}=\frac{B}{\mu}$ asymptotically, that is, suppose there exists some $\epsilon_{0}>0$ and $\delta (t_{0},\epsilon_{0})>0$, such that $||S(t_{0})-S^{*}_{0}||<\delta$, but $||S(t)-S^{*}_{0}||\geq \epsilon_{0}, \forall t\geq t_{0}$, then clearly from (\ref{ch1.sec2b.thm2.eq1}), either
\begin{equation}\label{ch1.sec2b.thm2.rem1.eq1}
 S^{*}_{0}=\lim_{t\rightarrow \infty}\frac{1}{t}\int_{t_{0}}^{t}S(\xi)d\xi\geq S^{*}_{0}+\epsilon_{0}\quad or\quad S^{*}_{0}=\lim_{t\rightarrow \infty}\frac{1}{t}\int_{t_{0}}^{t}S(\xi)d\xi\leq S^{*}_{0}-\epsilon_{0}.
\end{equation}
Thus, $\epsilon_{0}$ must be zero, otherwise  (\ref{ch1.sec2b.thm2.rem1.eq1}) is a contradiction. Hence, $E_{0}=(S^{*}_{0},0)$ is asymptotically stable.
\begin{rem}\label{ch1.sec2b.thm2.rem1}
Theorem~\ref{ch1.sec2b.thm2}, Theorem~\ref{ch1.sec2b.thm1}, [Theorem~3.1, Wanduku\cite{wanduku-biomath}] and Lemma~\ref{ch1.sec2b.lemma1} signify that  all trajectories of the solution  $(S(t), I(t)),t\geq t_{0}$ of the decoupled system containing equations for  $S$ and $I$ in the system  (\ref{ch1.sec0.eq3}) that start  in $D^{expl}(\infty)\subset D(\infty)$ remain bounded  in $D^{expl}(\infty)$. Moreover, the trajectories of the  infectious state $I(t), t\geq t_{0}$ of the solution $(S(t), I(t)),t\geq t_{0}$ in phase plane, ultimately turn to zero exponentially, whenever either the expected survival probability rate the malaria parasite satisfy  $E(e^{-(\mu_{v}T_{1}+\mu T_{2})})<\frac{1}{R^{*}_{0}}$, for $R^{*}_{0}\geq 1$, or whenever the basic production number satisfy $R^{*}_{0}<1$.  Furthermore, the rate of the exponential decrease of the infectious population from (\ref{ch1.sec2b.thm1.eq1}) is estimated by the  term $\lambda$, defined in (\ref{ch1.sec2b.thm1.eq1.proof.eq1.eq1}) and (\ref{ch1.sec2b.thm1.eq1.proof.eq2.eq1}).

In addition, Theorem~\ref{ch1.sec2b.thm2} asserts that when either the expected survival probability rate the malaria parasites satisfy  $E(e^{-(\mu_{v}T_{1}+\mu T_{2})})<\frac{1}{R^{*}_{0}}$, for $R^{*}_{0}\geq 1$, or whenever the basic production number satisfy $R^{*}_{0}<1$, the susceptible population remains strongly persistent in the mean over sufficiently large time, moreover, every trajectory of the susceptible state $S(t)$ that starts in $D^{expl}(\infty)$ remain bounded  in $D^{expl}(\infty)$, and  on average the trajectories converge to the disease free steady state population $S^{*}_{0}=\frac{B}{\mu}$.

From Theorem~\ref{ch1.sec2b.thm3}, since $(S(t), I(t))=E_{0}=(S^{*}_{0},0)=(\frac{B}{\mu},0)$ is the disease-free equilibrium of the decoupled system containing equations for  $S$ and $I$ in the system  (\ref{ch1.sec0.eq3}), and from (\ref{ch1.sec2b.thm2.eq1}), every trajectory for the susceptible state $S(t)$ converges asymptotically to $S^{*}_{0}=\frac{B}{\mu}$  on average, the disease-free steady state
$(S(t), I(t))\equiv E_{0}=(S^{*}_{0},0)=(\frac{B}{\mu},0)$ must be asymptotically stable.
%Indeed, if on the contrary, there exist a trajectory for $S(t)$ starting near $S^{*}_{0}=\frac{B}{\mu}$ that does not stay near $S^{*}_{0}=\frac{B}{\mu}$ asymptotically, that is, suppose there exists some $\epsilon_{0}>0$ and $\delta (t_{0},\epsilon_{0})>0$, such that $||S(t_{0})-S^{*}_{0}||<\delta$, but $||S(t)-S^{*}_{0}||\geq \epsilon_{0}, \forall t\geq t_{0}$, then clearly from (\ref{ch1.sec2b.thm2.eq1}), either
%\begin{equation}\label{ch1.sec2b.thm2.rem1.eq1}
% S^{*}_{0}=\lim_{t\rightarrow \infty}\frac{1}{t}\int_{t_{0}}^{t}S(\xi)d\xi\geq S^{*}_{0}+\epsilon_{0}\quad or\quad S^{*}_{0}=\lim_{t\rightarrow \infty}\frac{1}{t}\int_{t_{0}}^{t}S(\xi)d\xi\leq S^{*}_{0}-\epsilon_{0}.
%\end{equation}
%Thus, $\epsilon_{0}$ must be zero, otherwise  (\ref{ch1.sec2b.thm2.rem1.eq1}) is a contradiction. Hence, $E_{0}=(S^{*}_{0},0)$ is asymptotically stable.

In other words, over sufficiently long time, the population that remains will be all susceptible malaria-free people, and the population size will be equal to the disease free steady state population $S^{*}_{0}=\frac{B}{\mu}$ of the system (\ref{ch1.sec0.eq3}).% See $S^{*}_{0}=\frac{B}{\mu}$ in Wanduku\cite{wanduku-biomath}.
\end{rem}
 \section{Example for extinction of disease}\label{ch1.sec4}
%  It should be noted that some of the numerical examples discussed in this section are utilized in various capacities elsewhere to address different sub-objectives of the current on going project.
The examples exhibited in this section are used to facilitate understanding about the conditions for extinction of the disease in the population over time in Theorem~\ref{ch1.sec2b.thm2}. This objective is achieved in a simplistic manner by examining the behavior of the trajectories of  the different states ($S, E, I, R$) of the  system  (\ref{ch1.sec0.eq3}) over sufficiently long time.

%%%%
The following convenient list of parameter values in Table~\ref{ch1.sec4.table2} are used to generate and examine the trajectories of the different states of the system (\ref{ch1.sec0.eq3}), whenever the conditions of Theorem~\ref{ch1.sec2b.thm2} are satisfied.
  %%%%%
%  $R^{*}_{0}>1$, and the intensities of the white noise processes in the system continuously increase. It is easily seen that for this set of parameter values, $R^{*}_{0}=80.7854>1$. Furthermore, the endemic equilibrium for the system $E_{1}$ is given as follows:- $E_{1}=(S^{*}_{1},E^{*}_{1},I^{*}_{1})=(0.2286216,0.07157075,0.9929282)$.
  %in order to (1.) illustrate the impact of the source of the white noise processes in the system (owing to the random fluctuations in  the natural death or disease transmission rates) on the disease dynamics, and also to (2.) illustrate the effect of the intensity of the  white noise processes in the system on the trajectories of the different disease classes $(S, E, I, R)$ in the system. These illustrations also uncover the overall behavior of the stochastic system over time.
%%%%%%%%%%%
 %Figure~\ref{ch1.sec4.figure2} depicts the behavior of the stochastic system (\ref{ch1.sec0.eq8})-(\ref{ch1.sec0.eq11}) when the system has no fluctuations or whenever the intensity of the %fluctuations are infinitesimally small, that is $\sigma_{S}=\sigma_{E}=\sigma_{\beta}=\sigma_{I}=\sigma_{R}=0(\epsilon)$.
 \begin{table}[h]
  \centering
  \caption{A list of specific values chosen for the system parameters for the examples in subsection~\ref{ch1.sec4.subsec1}}\label{ch1.sec4.table2}
  \begin{tabular}{l l l}
  Disease transmission rate&$\beta$& 0.0006277\\\hline
  Constant Birth rate&$B$&$ \frac{22.39}{1000}$\\\hline
  Recovery rate& $\alpha$& 0.55067\\\hline
  Disease death rate& $d$& 0.011838\\\hline
  Natural death rate& $\mu$& $0.6$\\\hline
  %Intensity of fluctuations& $\sigma_{i}, i=S, E, I, R, \beta$& 0.05\\\hline
  Incubation delay time in vector& $T_{1}$& 2 units \\\hline
  Incubation delay time in host& $T_{2}$& 1 unit \\\hline
  Immunity delay time& $T_{3}$& 4 units\\\hline
  \end{tabular}
\end{table}
The Euler approximation scheme is used to generate trajectories for the different states $S(t), E(t), I(t), R(t)$ over the time interval $[0,T]$, where $T=\max(T_{1}+T_{2}, T_{3})=4$. The special nonlinear incidence  functions $G(I)=\frac{aI}{1+I}, a=0.05$ in \cite{gumel} is utilized to generate the numeric results. Furthermore, the following initial conditions are used
\begin{equation}\label{ch1.sec4.eq1}
\left\{
\begin{array}{l l}
S(t)= 10,\\
E(t)= 5,\\
I(t)= 6,\\
R(t)= 2,
\end{array}
\right.
\forall t\in [-T,0], T=\max(T_{1}+T_{2}, T_{3})=4.
\end{equation}
%%%%%%%%%%%%%%%%%infection-pop-extinction-sigma-SEIR-0-5-sigma-beta-2-5
%%%%%%%%%%%%%%%%
%%%%%%%%%%%%%%%%%
  %%%%%%%
  %%%%%%%%%%%%%%%%%
\begin{figure}[H]
\begin{center}
\includegraphics[height=6cm]{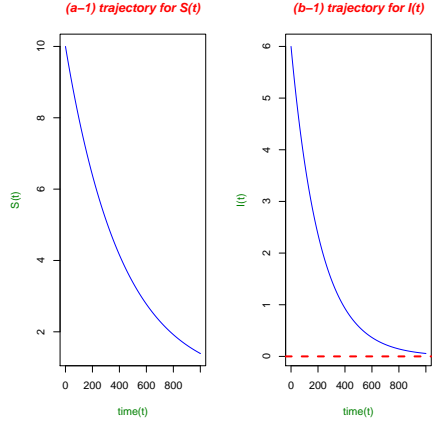}
\caption{(a-1), and (b-1),  show the trajectories of the  states $(S,I)$, respectively, over sufficiently long time $t=1000$, whenever the intensity of the incidence of malaria is $a=0.05$.
 %and the intensities of the white noise processes take the values $\sigma_{i}=0,\forall i\in\{S, E, I, R\}$, and $\sigma_{\beta}=10$.
%response intensity of malaria regulation is $\theta=1.5$. The broken lines represent the endemic equilibrium $E_{1}=(S^{*}_{1},E^{*}_{1},I^{*}_{1})=(0.247531,0.1074215,1.149901)$. Furthermore, $\min{(S(t))}=0.467897$, $\min{(E(t))}=-31.35931$, $\min{(I(t))}=6$ and $\min{(R(t))}=1.999997$.
Also, note that any  negative values have no meaningful interpretation, except a significance of extinction of the population if the negative values occur over sufficiently long time. Moreover, the basic reproduction number in (\ref{ch1.sec2.lemma2a.corrolary1.eq4}) in this case is $R^{*}_{0}= 2.014926e-05<1$, the estimate of the Lyapunov exponent or rate of extinction of the disease in (\ref{ch1.sec2b.thm1.eq1.proof.eq2.eq1}) is $\lambda= 1.162485>0$. The broken line in (b-1) signify the origin, while the broken line in (a-1) signify the disease-free-equilibrium $S^{*}_{0}=\frac{B}{\mu}=0.03731667$. The minimum value for $S(t)$ over time is $\min(S(t))=1.392842$, which is closer to $S^{*}_{0}=\frac{B}{\mu}=0.03731667$. The minimum value of $I(t)$ over time is $\min(I(t))=0.05678687$, near zero.
}\label{ch1.sec4.subsec1.fig2}
\end{center}
\end{figure}
%%%%%%%%%%%%%%%%%
%\begin{figure}[H]
   % %%%%%%%%%%%%%infection-pop-extinction-sigma-SEIR-0-0-sigma-beta-10
%   \begin{figure}[H]
%\begin{center}
%\includegraphics[height=6cm]{infection-pop-extinction-sigma-SEIR-0-0-sigma-beta-20.eps}
%\caption{(a-3), and (b-3),  show the trajectories of the  states $(S,I)$, respectively, over sufficiently long time $t=1000$, whenever the intensity of the incidence of malaria is $a=0.05$, and the intensities of the white noise processes take the values $\sigma_{i}=0,\forall i\in\{S, E, I, R\}$, and $\sigma_{\beta}=20$.
%%response intensity of malaria regulation is $\theta=1.5$. The broken lines represent the endemic equilibrium $E_{1}=(S^{*}_{1},E^{*}_{1},I^{*}_{1})=(0.247531,0.1074215,1.149901)$. Furthermore, $\min{(S(t))}=0.467897$, $\min{(E(t))}=-31.35931$, $\min{(I(t))}=6$ and $\min{(R(t))}=1.999997$.
%Also, note that all  negative values have no meaningful interpretation, except a significance of extinction of the population if the negative values occur over sufficiently long time. Moreover, the basic reproduction number in (\ref{ch1.sec2.lemma2a.corrolary1.eq4}) in this case is $R^{*}_{0}= 2.014926e-05<1$, the estimate of the Lyapunov exponent or rate of extinction of the disease in (\ref{ch1.sec2b.thm1.eq1.proof.eq2.eq1}) is $\lambda= 1.162485>0$. The broken line in (b-3) signify the origin, while the broken line in (a-3) signify the disease-free-equilibrium $S^{*}_{0}=\frac{B}{\mu}=0.03731667$.
%}\label{ch1.sec4.subsec1.fig3}
%\end{center}
%\end{figure}
%%%%%%%%%%%%%%%%%%

  Figure~\ref{ch1.sec4.subsec1.fig2} is used to verify the results about the extinction of the infectious population over time in Theorem~\ref{ch1.sec2b.thm1}, and the long-term behavior of the  susceptible population $S(t)$ in Theorem~\ref{ch1.sec2b.thm2}. Indeed, it can be observed that for the given parameter values in Table~\ref{ch1.sec4.table2}, and the initial conditions for the system  (\ref{ch1.sec0.eq3}) in (\ref{ch1.sec4.eq1}), it follows that the basic reproduction number in (\ref{ch1.sec2.lemma2a.corrolary1.eq4}) in this scenario is $R^{*}_{0}= 2.014926e-05<1$. Therefore, the condition of Theorem~\ref{ch1.sec2b.thm1}(a.) is satisfied, and from (\ref{ch1.sec2b.thm1.eq1.proof.eq2.eq1}), the estimate of the  Lyapunov exponent or the rate of extinction of the malaria population $I(t)$ is $\lambda= 1.162485>0$. That is,
\begin{equation}\label{ch1.sec4.subsec1.eq1}
  \limsup_{t\rightarrow \infty}{\frac{1}{t}\log{(I(t))}}\leq -\lambda = -1.162485.
\end{equation}
%\begin{equation}\label{ch1.sec2b.thm1.eq1.proof.eq2.eq1}
%\lambda\equiv(1-R^{*}_{0})(\mu+d+\alpha)>0.
%\end{equation}##1.80286
The Figure~\ref{ch1.sec4.subsec1.fig2}(b-1) confirms that over sufficiently large time, when $\lambda>0$, then the infectious population becomes extinct. Furthermore, note that the basic reproduction number in (\ref{ch1.sec2.lemma2a.corrolary1.eq4}) in this scenario is $R^{*}_{0}= 2.014926e-05<1$, which signifies that the disease is getting eradicated from the population over time, and the susceptible population seen in Figure~\ref{ch1.sec4.subsec1.fig2}(a-1)  over sufficiently long time, approaches the disease-free equilibrium state $S^{*}_{0}=\frac{B}{\mu}=0.03731667$.

Indeed, note that the minimum value for the trajectory of $S(t)$ in Figure~\ref{ch1.sec4.subsec1.fig3}(a-1) over long time is $\min(S(t))=1.392842$ near $S^{*}_{0}=\frac{B}{\mu}=0.03731667$. This suggests that the susceptible population is asymptotically stable on average over sufficiently long time near $S^{*}_{0}=\frac{B}{\mu}=0.03731667$ as shown in Theorem~\ref{ch1.sec2b.thm2}. Also, note that the general decrease in the susceptible population $S(t)$ in Figure~\ref{ch1.sec4.subsec1.fig2}(a-2) over time is accounted for  by the intensity of the incidence rate of malaria  $a=0.05$.
%\end{figure}
%%%%%%%%%%%%%%%%%
 \section{conclusion}
 The extinction and persistence of malaria in a family of SEIRS models is studied. Lyapunov functional, Lyapunov exponential analysis, and other analytic techniques are used to examine the trajectories of the system near the endemic and disease-free steady states of the system. The analytic results for extinction  dependent on the basic reproduction, and the expected probability of the survival of the malaria parasites. Moreover, numerical simulation results are presented to show the extinction of disease, and the asymptotic stability of the disease-free steady state population.

Also, from the above analysis, for the permanence of disease in the population, an extensive inherent algorithmic technique to analyze the permanence of disease in a complex multiple distributed delay system is presented in the proof of Lemma~\ref{ch1.sec4.lemma1}. The sufficient conditions for the permanence of disease are exhibited, and interpreted. Numerical simulation results are presented to investigate the impacts of malaria transmission rate and response to malaria regulation such as malaria treatment and improved living standards, on the permanence of the disease in the population.
%\newpage

\end{document}